\documentclass{emulateapj}
\usepackage{bm}
\usepackage{chngpage}
\usepackage{graphicx}
\usepackage[caption=false]{subfig}
\usepackage{mathrsfs}
\usepackage{threeparttable} 

\newcommand{\degree}{\ensuremath{^\circ}}
\def\apjl{ApJL}

\shorttitle{Jet Propagation}
\shortauthors{Geng et al.}

\begin{document}

\title{PROPAGATION OF RELATIVISTIC, HYDRODYNAMIC, INTERMITTENT JETS IN A ROTATING, COLLAPSING GRB PROGENITOR STAR}

\author{Jin-Jun Geng\altaffilmark{1, 2, 3}, Bing Zhang\altaffilmark{3}, Rolf Kuiper\altaffilmark{4}}

\altaffiltext{1}{School of Astronomy and Space Science, Nanjing University, Nanjing 210046, China; gengjinjun@gmail.com}
\altaffiltext{2}{Key Laboratory of Modern Astronomy and Astrophysics (Nanjing University), Ministry of Education, China}
\altaffiltext{3}{Department of Physics and Astronomy, University of Nevada Las Vegas, NV 89154, USA; zhang@physics.unlv.edu}
\altaffiltext{4}{Institute of Astronomy and Astrophysics, University of T\"ubingen, Auf der Morgenstelle 10, D-72076 T\"ubingen, Germany}

\begin{abstract}
The prompt emission of gamma-ray bursts (GRBs) is characterized by rapid variabilities, which may be a direct 
reflection of the unsteady central engine. 
We perform a series of axisymmetric 2.5-dimensional simulations
to study the propagation of relativistic, hydrodynamic, 
intermittent jets through the envelope of a GRB progenitor star.
A realistic rapidly rotating star is incorporated as the background of jet propagation, and the
star is allowed to collapse due to the gravity of the central black hole.
By modeling the intermittent jets with constant-luminosity pulses with equal on and off durations, we 
investigate how the half-period, $T$, affects the jet dynamics. 
For relatively small $T$ values (e.g. 0.2 s), the jet breakout time $t_{\rm bo}$ depends on the
opening angle of the jet, with narrower jets more penetrating and reaching the surface at
shorter times. For $T \leq 1$ s, the reverse shock crosses each pulse before the jet penetrates
through the stellar envelope. As a result, after the breakout of the first group of pulses at $t_{\rm bo}$,
several subsequent pulses vanish before penetrating the star, causing a quiescent gap. For larger
half-periods ($T=2.0, 4.0$ s), all the pulses can successfully penetrate through the envelope,
since each pulse can propagate through the star before the reverse shock crosses the shell.
Our results may interpret the existence of a weak precursor in some long GRBs, given that
the GRB central engine injects intermittent pulses with a half-period $T \leq 1$ s. The observational
data seem to be consistent with such a possibility.
\end{abstract}


\section{INTRODUCTION}
\label{sect:intro}

Gamma-ray bursts (GRBs) are extremely energetic explosions with enormous gamma-ray radiation. The
observational properties of GRBs demand that they originate from relativistic jets beaming towards earth 
(see \citealt{Piran04,Meszaros06,Kumar15} for reviews). 
Observations of GRB afterglows further reveal that these jets are narrowly beamed
with a typical opening angle around 0.1 rad \citep{Frail01,Liang08,Wang15}.
The so-called long GRBs (with durations longer than two seconds)
are widely believed to be associated with the death of massive stars \citep{Woosley93,Galama98,Stanek03,Hjorth03,Woosley06}.
The possible mechanisms to launch a relativistic jet from the central engine (likely a hyper-accreting black hole or
a millisecond magnetar) include magnetohydrodynamical processes (e.g., \citealt{Blandford77,Narayan12})
or neutrino annihilation processes (e.g., \citealt{Popham99,Kohri02,Gu06,Liu15}). After launching,
the jet propagates through and breaks out the stellar envelope of the progenitor star before emitting $\gamma$-ray
photons at a large radius. 

This paper focuses on the GRB jet propagation process inside the progenitor star. 
Previous studies have revealed some main characteristics of a propagating, constant-luminosity jet.
\cite{MacFadyen99} first used hydrodynamical simulations to show that a light jet can 
penetrate through a star and remains highly beamed. Later, jet propagation has been investigated by many 
authors using different codes that implement special relativity and improved numerical resolutions \citep[e.g.][]{Aloy00,ZhangW03,Mizuta06,Morsony07,Mizuta09,Mizuta13}.
These works showed detailed temporal and angular properties of the jets emerging from
massive stars. In the meantime, analytical studies have been carried out on the topic 
\citep{Waxman03,Matzner03,Bromberg11}.
According to these analytical and numerical approaches, the basic picture for jet propagation may be
summarized in the evolution of two phases
defined by the time domains before and after the jet breaks out the surface
of the star, respectively. During the first jet penetrating phase, one may spatially define five structures:
beam, envelope, jet head, cocoon, and multiple oblique collimation shocks.
When the beam is injected from the base of the star, a collimation shock \citep{Komissarov97,Komissarov98,Bromberg09,Mizuta13}
would emerge to generate the pressure needed to counterbalance the pressure from the surrounding
materials (the cocoon). The jet begins to propagate by pushing aside the matter in front of it. As a consequence,
a jet head consisting of a forward shock (FS) and a reverse shock (RS) would form.
The beam material shocked by the RS would expand due to high pressure, which forms
the jet cocoon. The stellar envelope materials shocked by the FS would also flow sideways
to form the jet cavity. The jet cocoon and jet cavity are called cocoon as a whole in this paper
for simplicity. 
During the jet's expedition to the stellar surface, some further collimating oblique shocks occur
due to a cycle of expansion and collimation of the jet \citep{Mizuta13}.
Meanwhile, the shear motion between the shocked beam and the shocked envelope materials
would produce some turbulent structures in the cocoon.
After the jet breaks out the star, the jet would be accelerated to a higher Lorentz factor
($\sim$ 5 times of the initial Lorentz factor) and the opening angle gets relatively smaller 
than the initial opening angle \citep{Morsony07,Mizuta13}.
Finally, the jet becomes internally free-expanding, bounded by a shear layer.

In most of these previous simulations, a steady, constant-luminosity has been assumed. In reality, GRB jets
are likely intermittent in nature, as revealed by their erratic lightcurves \citep{Fishman95}.
Even though the emission sites and processes that give rise to the observed GRB lightcurves
are still subject to debate (e.g. photospheric emission \citep{Paczynski86,Rees05,Lazzati13,Peer16},
internal shocks \citep{Rees94,Kobayashi97}, or internal collision-induced magnetic reconnection
and turbulence \citep{Zhang11,Zhang14,Deng15}), all these models attribute all or part of the variability time
scales to the intrinsic intermittent activities \citep[e.g.][]{Fenimore99,Yuan12}
or the precession activities \citep[e.g.][]{Ito15} at the central engine. 

Numerical studies of propagation of intermittent jets in GRB progenitor stars have been carried out
by \cite{Morsony10,Lopez14,Lopez16}. They showed that the fast engine variability in the injected 
jets could be preserved after the jet emerges from the star.
\cite{Lopez16} showed that the interaction between the episodic jet and the
progenitor stellar envelope can lead to an asymmetric behavior in the light curves of GRBs.
Their results suggested a correlation between the duty cycle period of the central engine variability and the breakout
time. Furthermore, they pointed out that the general trend and behavior in the three-dimensional (3D) simulations
are basically similar to those in the two-dimensional (2D) simulations.
However, the envelope model used in their simulations is simplified, with the progenitor star as
a background without rotation or infall due to core collapse.

In this paper, we investigate the propagation of relativistic, hydrodynamic, intermittent GRB jets in
a star with more realistic physical conditions through numerical simulations.
The progenitor star used in our simulations is rotating and collapsing due to gravity.
In order to find out how intermittent jets drill out the star, we perform a series of 2.5-dimensional (2.5D, 2D 
plus rotation) simulations in which jets are injected intermittently from the central engine.
For simplicity, the intermittent jets are characterized as a periodic step function with the same active and 
quiescent durations, $T$, which is half of the duty cycle period.
From these simulations, we intend to study how the parameter $T$ affects the ability of
the jet to break out the star.

This paper is organized as follows. Numerical simulation setup is presented in Section 2.
Detailed information for the numerical methods is described in Section 3.
In Section 4, we show the numerical results, which are followed by a discussion of the analytical
understanding of these results in Section 5.  
The conclusions are summarized in Section 6.

\section{SIMULATION SETUP}

\subsection{The Envelope of a Rotating Star}

Our first step is to obtain the structure of a rotating massive star envelope in dynamical
equilibrium. According to \cite{WoosleyH06}, the 16TI model is one of the promising GRB progenitor models.
In this subsection, we try to construct our envelope model to mimic the 16TI model.

Since the envelope of 16TI model is nearly radiation-dominated, a polytropic equation
of state $p = K \rho^{4/3}$ is adopted, where $p$ is pressure and $\rho$ is density.
We consider a simple barotropic star, i.e., $K$ is a constant.
After assuming that the star is axisymmetric about its rotational axis, the main task left is 
to solve the Euler equations with rotation (see \citealt{Fujisawa15})
\begin{eqnarray}
\frac{1}{\rho} \frac{\partial p}{\partial r} = - \frac{\partial \phi}{\partial r} + r \sin^2 \theta \Omega^2, \\
\frac{1}{\rho} \frac{\partial p}{\partial \theta} = - \frac{\partial \phi}{\partial \theta} + r^2 \sin \theta \cos \theta \Omega^2,
\end{eqnarray}
and the integral form of the Poisson equation
\begin{equation}
\phi = -G \int \frac{\rho (\bm{r}^{\prime})}{|\bm{r} - \bm{r}^{\prime}|} d^3 \bm{r}^{\prime}.
\end{equation}
Here $p$, $\rho$, the gravitational potential $\phi$ and angular velocity $\Omega$ are all functions of
$r$ and $\theta$, and $G$ is the gravitational constant. Spherical polar coordinates ($r$, $\theta$) are
adopted in the formulation.

Before solving Equations (1-3), a one-dimensional (1D) spherical equilibrium non-rotating star
is used as the initial guess for the 2D rotating star.
We first solve the simple equation 
\begin{equation}
\frac{d p}{d r} = -\frac{G m \rho}{r^2},
\end{equation} 
where $m$ is the enclosed mass.
We integrate this equation from a starting radius $r_s$ ($10^6$ cm), using the Runge-Kutta method.
The mass inside $r_s$ is set as $1.7 M_{\odot}$, and the density at $r_s$
is set as $6.3 \times 10^9$ g~cm$^{-3}$, so that the radius of this 1D star
is $\sim 4 \times 10^{10}$ cm, and the total mass of the star is $\sim 14 M_{\odot}$. 
Using this 1D solution as an initial guess, we then apply the Hachisu's Self-Consistent Field (HSCF) method
(see \citealt{Hachisu86a,Hachisu86b,Fujisawa15} for details) to get the 2D configuration of 
a rotating star. A realistic star may have different layers obeying different rotation laws due to differential rotation \citep{Kiuchi10}.
Detailed simulations have shown that the distribution of the angular velocity as a function of the mass coordinate 
may be approximated as a step-function \citep{Maeder12}. The angular velocity may
be treated as a constant in the core region that encloses a mass $M\leq 7 M_{\odot}$ 
(\citealt{Hirschi04}, see also Figure 16 of \citealt{Maeder12}).
Since our problem mostly concerns the core region of the star where the jet penetrates, for simplicity, 
we assume rigid rotation, i.e., $\Omega=$constant, throughout the paper to ease the numerical simulation problem.
Moreover, the axis ratio $q = R_{\rm pol} / R_{\rm eq}$
($R_{\rm pol}$ and $R_{\rm eq}$ are the polar radius and the equatorial radius respectively) is taken as 0.7,
which corresponds to a critical (or break-up) rotation configuration \citep{Maeder12}.
Our results give $\Omega = 5.3 \times 10^{-3}$ rad~s$^{-1}$, which is consistent with the value in \citep{Maeder12} and
\cite{Nagakura11}\footnote{See also Nagakura's doctor thesis, http://dspace.wul.waseda.ac.jp/dspace/handle/2065/37618}.
One may note that the surface velocity of this star is larger than observed (e.g., \citealt{Dufton11}),
which is mainly due to the assumption of rigid rotation.
However, this would not strongly affect the results of jet propagation study below since
the jet is moving along the polar direction.
The upper panel of Figure 1 shows the density distribution of the rotating star, which is used as an input in the next step.

\subsection{Collapse of the Star}

According to the popular GRB central engine models, the GRB jet is launched through the accretion of 
envelope materials into the central black hole (BH). Therefore as the jet is launched, the star
should undergo collapsing after the core collapses to a BH and losses radiation support.
In the following, we try to set up a collapsing star into which the GRB jet is injected.

In our simulations, we treat the inner region ($r \leq 4 \times 10^8$ cm) of the star as a newly
formed BH, of which the mass is $\sim 2.0 M_{\odot}$.
The collapsing of the envelope can be modeled by setting the boundary condition to be
a type of outflow (the gradients of all quantities are zero, also see \citealt{Nagakura11}).
The rarefaction wave triggered at the inner boundary would cause infall of the matter
at the radius it reaches. 
Since the self-gravity module in the PLUTO code (see Section 3 for details) is not
fully implemented yet\footnote{A
self-gravity module for PLUTO has been implemented for specific problems before, see \cite{Kuiper10,Kuiper11}},
we adopt an approximation method to treat the effects of gravity.
For a point at radius $r$, its acceleration in the radial direction is calculated as
\begin{equation}
g = - \frac{G M_r}{r^2},
\end{equation}
where $M_r = 4 \pi \int_{0}^{\pi/2} \int_{r_0}^{r} \rho(r^{\prime}) r^{\prime 2} \sin \theta dr^{\prime} d\theta + M_{\rm BH}$,
$r_0 = 4 \times 10^8$ cm is the inner boundary, and $M_{\rm BH}$ is the mass of the central BH.
Notice that $M_{\rm BH}$ is increasing with time due to the infall of matter in our calculation.
The justification of this treatment can be seen in the lower panel of Figure 1, where we plot
the specific angular momentum $j$ at the innermost stable circular orbit (ISCO) for both a Schwarzschild 
black hole (green) and a Kerr black hole (red) as a function of the black hole mass, 
together with the equatorial specific angular momentum $j$ of the star (blue) as a function of 
the mass enclosed. During the jet launching phase, the black hole in our simulation grows from
$2 M_\sun$ to $\sim 6 M_\sun$. One can see that at the equator, the stellar $j$ is larger than
$j$ at ISCO, so that infall is impossible. This is consistent with the formation of an accretion torus
that powers the GRB \citep{WoosleyH06}. On the other hand, the $j$ values of the star decreases with
decreasing polar angle $\theta$, so that below a critical angle,  vertical infall into the black hole
becomes possible. At the extreme case with $\theta=0$, $j$ is essentially zero for the stellar materials.
In our simulation, when gravity is turned on, steady growth of the black hole is observed.

Using the envelope solution obtained in Section 2.1, we simulate the collapsing process using the
hydrodynamic module provided in PLUTO and get the structure of the star at a specific time.
We then move on to launch a relativistic jet from $r_0$ starting from this epoch.
When the jet is launched, we neglect self-gravity of the envelope, but consider the point-mass gravity
of the central BH.

\subsection{Jet Propagation}

Since our objective is to study jet propagation, we simply inject a relativistic jet via boundary conditions
without simulating the jet launching processes \citep[e.g.][]{Nagataki07,Nagataki09} from the first principles.
The jet is characterized by three parameters, i.e., luminosity ($L_j$),
initial Lorentz factor ($\Gamma_0$, corresponding velocity $\beta_0 c$), and half-opening angle ($\theta_0$).
The boundary conditions can be set up correspondingly. 
We denote density, pressure, and internal energy density of the jet in the fluid frame as 
$\rho_j$, $p_j$, and $e_j$ respectively,
then the energy density in the lab frame can be expressed as
\begin{equation}
e_{\rm lab} = \Gamma_0^2 (\rho_j c^2 + e_j + \beta_0^2 p_j) \simeq \Gamma_0^2 \left(1 + \frac{4 p_j}{\rho_j c^2} \right) \rho_j c^2,
\end{equation}
where we have assumed the adiabatic index as 4/3 ($e_j = 3 p_j$), and $\beta_0 \approx 1$.
On the other hand, $e_{\rm lab}$ is related to $L_j$ by
\begin{equation}
e_{\rm lab} = \frac{L_j}{2 \pi r_0^2 (1-\cos \theta_0) \beta_0 c}.
\end{equation}

When the internal energy of the material is fully converted to kinetic energy,
it would reach a terminal Lorentz factor $\Gamma_\infty$, which is calculated as
\begin{equation}
\Gamma_\infty = \left(1 + \frac{4 p_j}{\rho_j c^2} \right) \Gamma_0.
\end{equation}
Equations (6-8) are essential to infer $\rho_j$, $p_j$ (initial conditions of the jet) 
given $L_j$, $\theta_0$, $\Gamma_0$, and $\Gamma_\infty$.

In this paper, we fix $L_j = 5 \times 10^{50}$ erg~s$^{-1}$, $\Gamma_0 = 5$,
and $\Gamma_\infty = 400$, unless noted otherwise.
The initial opening angle $\theta_0$ is set to vary to investigate its effect on jet propagation.

\section{NUMERICAL METHODS}
All the simulations presented here were performed using the PLUTO code, version 4.2.
The PLUTO code is built on the Godunov-type shock-capturing schemes, aiming to the solution of 
Newtonian, relativistic flows (see \citealt{Mignone07} for a full description).

In this work, spherical coordinates ($r$, $\theta$) are employed and axisymmetry is assumed for all simulations.
The collapsing process of the star is simulated using the hydrodynamic (HD) module provided in PLUTO.
The star is immersed in a medium with a constant density $10^{-10}$~g~cm$^{-3}$.
The computational domain covers a region of $4 \times 10^8~\mathrm{cm} \leq r \leq 1.0 \times 10^{11}~\mathrm{cm}$
and $0\degree \leq \theta \leq 90\degree$.
The radial grid consists of 2048 points and is logarithmically distributed,
while the angular grid is uniform with 256 points.
A Riemann solver, called HLL solver (Harten-Lax-Van Leer approximate Riemann solver, see \citealt{Harten83}),
a linear-type spatial reconstruction, and a second-order Runge-Kutta time integration
were chosen in the simulations.

Jet propagation is simulated using the relativistic hydrodynamic (RHD) module in PLUTO.
Specifically, we choose an extended Harten-Lax-Van Leer Riemann solver (called HLLC solver, see \citealt{Harten83,MigBodo05}),
a so-called ``OSPRE\_LIM'' flux limiter using a piecewise linear interpolation \citep{Mignone05},
and a second-order Runge-Kutta time integration. 
As a result, we achieve second-order accuracy in both space and time.
The computational domain covers a region of $4 \times 10^8~\mathrm{cm} \leq r \leq 1.0 \times 10^{11}~\mathrm{cm}$
and $0\degree \leq \theta \leq 90\degree$.
Adaptive mesh refinement \citep{Mignone12} is adopted to improve the computational efficiency.
There are altogether 6 grid levels in our setup, of which the base grid consists of $128 \times 64$ uniform grid cells,
and the refinement ratio is 2. 
With this setup, the highest resolution of the grid is $(\Delta r, \Delta \theta) = (2.4 \times 10^7 \rm{cm}, 0.04\degree)$,
which is comparable with previous 2D studies (\citealt{Morsony07,Mizuta13}).

\section{SIMULATION RESULTS}

In order to ensure that our simulation results are reliable, we first
perform one baseline simulation to directly compare with the results of \cite{Morsony07}.
In this simulation, we use a simple power-law model for a non-rotating star with a mass of $15 M_{\odot}$ and
a surface radius of $10^{11}$ cm, without considering gravity.
All the jet parameters are the same as those of \cite{Morsony07},
i.e., $L_j = 5.32 \times 10^{50} \mathrm{erg}~\mathrm{s}^{-1}$, $\Gamma_{\infty} = 400$,
and $\theta_0 = 10\degree$.
The inner boundary is set to $r = 10^9$ cm and the grid setup is the same as that in Section 3.
Figure 2 upper panel shows the density distributions at three epochs, $t=9.0, 13.0, 16.0$ s, respectively.
It can be noticed that the jet breakout time is $\sim 13$ s,
which is almost the same as the value of \cite{Morsony07}.
Moreover, the structure of the collimation shock, the cocoon, and the bow shock
after the jet breakout time are also in good agreement with previous works.
On the other hand,
low-resolution spherical coordinates would lead to numerical baryon loading \citep{Mizuta13}.
We then diagnosed the resolution of our setup by checking whether Bernoulli's constant $h \Gamma$ 
($h = 1 + e_j/(\rho_j c^2) + p_j/(\rho_j c^2)$ is the specific enthalpy) 
along the jet axis is conserved.
Figure 3 shows that $h \Gamma$ along the jet axis is conserved up to the jet head,
except for some fluctuations due to internal shocks.
Also in this figure, the evolution of $\Gamma$ shows that the core of the 
unshocked jet is free-streaming up to the converging position of the collimation shock.
Therefore, using PLUTO our grid setup is reliable for the purpose of this research. 

The star is already collapsing before jet launching.
However, the exact jet launching delay time is not known.
We choose a moderate value for this delay time, i.e., 24 s, in our simulations.
Figure 4 shows the density profiles in the equatorial and axial directions for $t=0$ and $t=24$ s
after the collapse, respectively.
A shock wave produced by centrifugal bounce is predicted to emerge during the
collapse \citep{Nagakura11}. This shock wave does not appear in Figure 4.
This may be caused by the approximate treatment used here on self-gravity (Equation 5).
However, this would not affect much our following results, since the
primary density profile in Figure 4 is still similar to that in \cite{Nagakura11}.
Using this envelope as the initial condition, we perform two series of simulations for intermittent jets.
The jets are characterized by periodic injections with top-hat constant luminosity in each episode
and with same duration during the on and off states. We simulate four intermittent jets with half period
$T=0.2$, 1.0, 2.0, and 4.0 s, respectively. For comparison, we also simulate a baseline constant luminosity jet. 
For each case, we perform two simulations with the jet opening angle being $5^{\rm o}$ and $10^{\rm o}$,
respectively. 
The simulation models are named in the form of ``mN1TN2'', with
N1 denoting the value of the $\theta_0$ and N2 denoting the value of $T$. For constant luminosity
jets, N2 is replaced by ``con''. The detailed parameters for all these models are listed in Table 1.
The density distributions of the m10Tcon model at three epochs, $t=3.0, 4.4, 6.2$ s, respectively, 
are presented in the lower panel of Figure 2.

From these simulations, we investigate the relationship between the breakout time ($t_{\rm bo}$)
and the half period ($T$), which is presented in Figure 6.
It is interesting to note that at $T = 0.2$ s, $t_{\rm bo}$ is significantly different for the two different $\theta_0$
values, while for other values of $T$, $t_{\rm bo}$ depends weakly on $\theta_0$.
In comparison with the results of \cite{Lopez16}, under the same initial condition ($\theta_0 = 10 \degree$), 
the trend of our $t_{\rm bo}-T$ relation is similar to the 3D result in \cite{Lopez16}, but is 
different from their 2D result. 
We should note that the oblate configuration of the envelope is the main reason that
makes the difference of the absolute values of $t_{\rm bo}$ between our results
and the results in \cite{Lopez16}.

Figures 7 and 8 show the snapshot Lorentz factor maps (color) and density distributions (grey) for two sets of 
intermittent jet simulations. The cases of constant luminosity jets are presented in Figure 5 for comparison.
The epoch for the snapshot for each model roughly corresponds to the 
time when the first episode of the jet reaches the simulation boundary. 
To identify fast moving ejecta, only regions with $\Gamma > 2$ are shown in color. 
From Figs. 7 and 8, one can see spatially separated high-$\Gamma$ regions that manifest the
intermittent injection of the central engine.

One most striking finding from our simulations is that intermittent jets with a relatively short half-period
do not have every sub-pulse emerges from the star. After the first pulses escape the star at $t_{\rm bo}$,
some sub-pulses vanish during their propagation inside the star. Here we define ``jet vanishing" as 
Lorentz factor drops below 2. 
For example, for model m10T0.2, after $t_{\rm bo}$ (6.47 s), the next jet pulse that can reach the outer 
boundary breaks out the star surface at $\sim$ 11.6 s.
For model m10T1.0, after $t_{\rm bo}$ (6.0 s), only the 7th pulse (ejected during 12--13 s) and later ones
are able to reach the outer boundary. Figure 9 gives some details of m10T1.0 simulation. The top panel
($t=7.2$ s) shows that the first three pulses (p1-3) merge and break out from the star at $Y \geq 150$, and 
that the 4th pulse (p4) was just ejected from the engine (head at $Y \geq 40$). In the next two panels ($t=8.2, 9.2$ s),
one can see p4 gradually vanishes. The fifth pulse (p5) seen in the $t=9.2$ s panel and the sixth pulse
(p6) seen in the $t=11.2$ s panel all gradually vanish before penetrating through the star. Only until pulse 7
(p7), as seen in the $t=13.2$ s panel manages to breakout the star again at $t=15.6$ s.

For relatively large half-period intermittent jets ($T =2, 4$ s), the sub-pulse vanishing events do not happen. 
Every sub-pulse can manage to break out the star separately.

\section{ANALYTICAL UNDERSTANDING}

The numerical results can be understood using some analytical treatments.

Let us first consider the propagation of a continuous jet. 
Our numerical results (m5Tcon and m10Tcon) can be directly compared with the results of the analytical model
\citep[e.g.][]{Bromberg11}.
The jet head velocity is given by
\begin{equation}
\beta_h = \frac{\beta_j}{1+\tilde{L}^{-1/2}},
\end{equation}
where $\tilde{L} = \frac{L_j}{\Sigma_j \rho_a c^3}$, $\rho_a (r,\theta=0)$ is
the ambient medium density in the polar direction,
$\Sigma_j$ is the jet cross section and can be further derived by considering 
the pressure balance between the jet and the cocoon. 
In order to get the analytical formulae for $\Sigma_j$, four coefficients, $\xi_a$, $\xi_h$, $\xi_c$, and $\eta$, 
are introduced to simplify the derivation (see \citealt{Bromberg11,Mizuta13} for details). 
In short, $\eta$ is a parameter to correct the approximation of the cylindrical cocoon shape,
$\xi_a$ represents the correction to the mean density of medium in the cocoon,
$\xi_h$ and $\xi_c$ are the corrections to the approximations of jet head position and cocoon width, respectively.
When the density profile in the poloidal direction is a power-law $\rho_a \propto z^{-\alpha}$,
$\xi_a$, $\xi_h$, $\xi_c$ can be expressed as $\xi_a = \frac{3}{3-\alpha}$, $\xi_h = \xi_c = \frac{5-\alpha}{3}$.
After introducing these coefficients, 
the jet head velocity and jet head position can be finally calculated as
\begin{equation}
\beta_h \simeq \tilde{L}^{1/2} \simeq \left( \frac{L_j}{c^5 t^2 \rho_a \theta_0^4} \right)^{1/5} 
\left( \frac{16}{3 \pi} \frac{\eta \xi_a}{\xi_h \xi_c^2} \right)^{1/5},
\end{equation}
\begin{eqnarray}
z_h & = & \xi_h \beta_h c t \simeq \left( \frac{t^3 L_j}{\rho_a \theta_0^4} \right)^{1/5}
\left( \frac{16}{3 \pi} \frac{\eta \xi_a \xi_h^4}{\xi_c^2} \right)^{1/5}   \nonumber \\
& \simeq &1.4 \times10^{10} \mathrm{cm} \left( \frac{t}{1 \mathrm{s}} \right)^{3/5}  \left( \frac{L_j}{10^{51} \mathrm{erg}~\mathrm{s}^{-1}} \right)^{1/5}  \nonumber \\
& \times & \left( \frac{\rho_a}{10^3 \mathrm{g}~\mathrm{cm}^{-3}} \right)^{-1/5}  \left( \frac{\theta_0}{0.1} \right)^{-4/5} \left( \frac{\eta}{0.01} \right)^{1/5},
\end{eqnarray}
where $\alpha = 2$ is assumed (which gives $\xi_a = 3$, $\xi_h = \xi_c = 1$). 
By adjusting the parameter $\eta$, we can fit the numerical results
with the analytical solution in Equation (11).
Figure 10 shows the comparison between the analytical solution ($\eta = 0.02$ for $\theta_0 = 5\degree$
and $\eta = 0.006$ for $\theta_0 = 10\degree$) and the numerical results.
The numerical results are in good agreement with the analytical results.
In this figure, the analytical solutions are obtained using Equation (11) with the density slope
$\alpha = -d\mathrm{ln} \rho_a /d\mathrm{ln} z$ at $z = z_h$.

For $T=0.2$ s intermittent jets, $t_{\rm bo}$ in m5T0.2 and m10T0.2 are significantly different from each other.
The narrow jet (m5T0.2) travels much faster ($t_{\rm bo} = 4.86$ s) than the wide jet (m10T0.2)
($t_{\rm bo}=6.47$ s). This is caused by two reasons. First, 
for a smaller $\theta_0$, the energy is deposited on a smaller area so that the jet is more penetrating
(jet pulse is narrower and move faster) in the polar direction.
Second, the breakout in these short $T$ models is triggered by the expanding cocoon.
In these models, stellar materials are shocked into a hot cocoon matter by the vanishing jet pulses.
When a bow shock forms at the stellar surface, there is only a small amount of jet matter (with high 
$\Gamma$) near the surface, so that the outgoing materials are mostly from the cocoon.
The larger the $\theta_0$, the wider the cocoon (see Figure 11), and the smaller the expanding
velocity in the polar direction.

For longer duty cycle jets, i.e. $T \geq 1.0$~s, $t_{\rm bo}$ is not sensitive to $\theta_0$.
This is because the breakout of these cases is 
triggered by the jet pulses rather than the cocoon. As a result, given the same $T$, 
the jet breakout times for different $\theta_0$ values are defined by the same ejection
time from the central engine.
For example, in m5T2.0 and m10T2.0, the 2nd jet pulse ejected during 4--6 s is
powerful enough to lead to the breakout in both cases.

The jet pulse vanishing in m10T0.2 and m10T1.0 but not in m10T2.0 and m10T4.0 
can be understood.
Like a continuous jet, a jet head is also formed to evacuate the envelope matter in front of the jet pulse.
The velocity of the jet head is smaller than that of the jet itself,
so that the jet materials would flow across the RS into the jet head, and finally flow into the cocoon.
If the jet pulse does not last for a long time, all the jet materials would
go into the cocoon before the breakout of the jet head.
For a jet pulse of an active duration of $T$, we assume that the average velocity
of the jet head is $\bar{\beta}_h$. It would vanish
after a duration 
\begin{equation}
\tau \simeq \frac{\bar{\beta}_h T}{1-\bar{\beta}_h} = \frac{T}{1/\bar{\beta}_h - 1}.
\end{equation}
By equaling $\bar{\beta}_h  c \tau$ and $R_{\rm pol}$,
one can obtain a critical duration 
\begin{equation}
T_c = \frac{1}{\bar{\beta}_h } (\frac{1}{\bar{\beta}_h } - 1) \frac{R_{\rm pol}}{c},
\end{equation}
below which the jet crosses the jet head before penetrating through the star, so that
it would vanish. 
In our calculations, $R_{\rm pol} \simeq 2.8 \times 10^{10}$ cm is adopted.
Our simulations show that $\bar{\beta}_h \simeq 0.6$ could be reached for each pulse.
One then gets $T_c \sim 1.0$ s. As a result, a jet pulse would be able to drill out the
stellar surface and reach the outer boundary if $T > T_c \sim 1.0$ s is satisfied.
This is consistent with the fact that jet vanishing does not appear in m5T2.0 or m10T2.0.
Another condition for a successful jet pulse is that the next jet would not catch up with
the prior jet within the duration $\tau$, so a quiescent duration should not exceed 
$(1-\bar{\beta}_h) \tau$. 
Both conditions are satisfied for the $T=2.0, 4.0$ s cases, but not satisfied for
$T=0.2$ s and $T=1$ s cases (see Figure 12). This explains the jet pulse vanishing phenomenon in our 
simulations with small $T$.

The collapsing envelope also plays a key role in destroying jet pulses.
In comparison with m10T0.2,
we run a simulation in which the star is not rotating or collapsing, and the jet
is launched at $r_0 = 10^9$ cm (a common value used by other authors).
It is found that the jet vanishing does not exist in this simulation (see Figure 13).
This difference also holds if we move the injection radius in m10T0.2 
from $4 \times 10^8$~cm to $10^9$~cm.
In Figure 13, we can see that the cocoon structure without gravity is obviously
wider than that in m10T0.2. This in turn gives a larger pressure in the hot cocoon,
which makes narrower jet pulses and stronger internal shocks to dissipation jet energy
(Figure 14).
Even though gravity does not affect jet dynamics directly, our results suggest
that gravity in addition to rotation may influence the cocoon structure, and hence the jet dynamics.

The jet pulse vanishing phenomenon in m10T0.2 or m10T1.0 indicates that
there may be a time delay between the first breakout and the 
breakouts of subsequent jet pulses.
In the case of m10T1.0, the time delay can reach $\sim$ 7 s.
This value could be even larger, if the engine quiescent time duration is even longer than the
active time (in our simulation we assumed the same duration). 
This motivates us to relate our simulation results to the lightcurve quiescent time between
the GRB precursor and its main emission.
It has been found that $\sim$15\% of GRBs have a precursor \citep{Burlon08,Burlon09}.
The mean duration of the quiescence is $\sim$ 50 s, of which the origin
is still uncertain \citep{Meszaros00,Ramirez02,Lazzati05,Wang07,Hu14}.
Our results can provide a reasonable explanation for such a quiescent time.
According to our picture, the GRB central engine may continuously eject episodic
jets with short active time scales. Initially, the several leading sub-pulses merge and
break out the star with the mixed cocoon and jet matter, which produces the
precursor. Later, for a certain duration, the episodic jet pulses injected from the
central engine gets vanished since the vanishing time $\tau$ is shorter than the
jet penetrating time. The GRB therefore displays a long gap of quiescence.
Finally, the funnel near the jet axis becomes more and more hollow,  
some episodic jets of long active periods can eventually escape the star to make radiation,
making the main episode of the burst. 
According to our simulations, pulse vanishing happens only if
$T$ is around or shorter than 1 s. Observationally, the shortest variability time scale for
long GRBs is of the similar order. This also lends support to our model.
Recently, \cite{Lloyd16} showed that within the context of a magnetically arrested disk around a BH,
the characteristic timescale for the variability of the Poynting flux jet is $\sim 1$~s.
This also provides a physical standpoint for our scenario, i.e., the jet pulses are characterized by a duty
cycle period of the order of a second.
However, we have not considered the role of magnetic fields on jet propagation in this work.

There are two facts that support our proposal.
The precursors and main events have very similar spectral properties \citep{Burlon09,Hu14},
which implies that they share the similar radiation physics.
In our explanation, the precursor and the later main bursts are all from episodic jet pulses
ejected from the same central engine, so that their emission properties should be similar.
Second, multiple precursors have been observed in many cases (e.g. 73 cases in the \cite{Burlon09} sample).
This fact is also a natural outcome in our scenario. After the initial breakout to produce the first precursor, 
in some GRBs, the episodic jets may have relatively long active durations, so that they can subsequently 
break out the star to produce more precursors before the main burst comes. Some others may have 
relatively short active time durations, so that they can only produce one precursor with a long quiescent 
gap before the main burst comes after the funnel is cleared. 

Another interesting finding in our simulations is that later injected pulses are likely to catch up with the
pulses injected earlier. GRB internal shock models require that late shells to catch up with the early ones.
In our simulations, even if all the pulses have the same initial energy and internal energy, later injected
jet pulses are accelerated more quickly when
the episodic jets propagate in the collapsing envelope.
This may be understood as follows. 
First group of pulses that break out the star at $t_{\rm bo}$ and make the precursor would evacuate
a funnel along the jet axis direction, which cannot be refilled quickly.
As a consequence, pulses injected at later epochs would travel in a more stratified medium and therefore
can be accelerated more quickly.
In m5T0.2, later pulses can even catch up with the early ones within the simulation box  (see Figure 15).

\section{CONCLUSIONS AND DISCUSSION}

In this paper, we numerically modeled the propagation of a relativistic, hydrodynamic, intermittent jet 
in the envelope of a massive, rotating, and collapsing star for the first time.
Using the PLUTO code with specific setups, we performed a set of 2.5D simulations. 
Because of the introduction of a more realistic rotating star and central gravity during the jet
propagation phase, the following interesting results (some of which were not obtained previously)
are obtained:

For intermittent jets, the shock breakout time $t_{\rm bo}$ depends on the jet opening angle when
the half-period $T$ is short enough (e.g. $T = 0.2$~s). This is because for short $T$, the reverse
shock to each pulse crosses the shell before it penetrates through the star, so that the shock breakout
is defined by that of a hot cocoon, the speed of which depends on the opening angle of the jet.
When the half period is large enough ($T \geq 1$ s), the reverse shock crossing time becomes comparable
or longer than the jet penetrating time, so that $t_{\rm bo}$ no longer sensitively depends on the jet 
opening angle. The $t_{\rm bo}-T$ relation trend obtained from our 2.5D results, is similar to
that from 3D results in \cite{Lopez16}. 

For short half-period jets ($T=0.2, 1$ s), our simulations revealed some vanishing pulses after
the first group of pulses break out the star. The vanish of sub-pulses is due to two reasons: that
the reverse shock crosses the pulse before it penetrates the star and that a later pulse may have caught up
with a preceding jet. Such vanishing effect gives rise to a quiescent time after the initial shock
breakout even if the engine is continuously ejecting intermittent pulses, and can be a mechanism
to interpret the quiescent gap between the precursor and the main burst. Such a behavior is 
only observed when a rotating, collapsing star is simulated. 

Growing evidence suggests that GRB prompt emission jets may
carry a significant Poynting flux
\citep[e.g.][]{Zhang13,Uhm14,Uhm16}.
In the future, we plan to introduce strong magnetization in relativistic jets and
numerically investigate the effect of strong magnetization on jet propagation.
A more self-consistent setup of the collapsar including the magnetic fields 
may be also considered \citep{Mosta14,Mosta15}.

\acknowledgments
We thank the anonymous referee for valuable suggestions, and 
Kotaro Fujisawa, Davide Lazzati, Diego L{\'o}pez-C{\'a}mara, Andrea Mignone, 
Brian Morsony, Philipp M{\"o}sta, and Shigehiro Nagataki for helpful comments on the paper.
This work is partially supported by NASA under grants NNX15AK85G and NNX14AF85G,
the National Basic Research Program of China with Grant No. 2014CB845800, 
and by the National Natural Science Foundation of China (grant Nos. 11473012 and 11303013).
J.J.G. acknowledges the China Scholarship Program to conduct research at UNLV.
R.K. acknowledges financial support within the Emmy Noether research group 
on ``Accretion Flows and Feedback in Realistic Models of Massive Star Formation''
funded by the German Research Foundation under grant no. KU 2849/3-1.
The visualization of the simulation data in this article is conducted with software VisIt \citep{Childs05}.
This work made use of Cherry Creek cluster at UNLV's National Supercomputing Institute.

\newpage

\begin{deluxetable}{cccccccc}
\tabletypesize{\scriptsize}
\tablewidth{0pt}
\tablecaption{Initial Conditions of the Episodic Jet Models\label{TABLE:1}}
\tablehead{%
        \colhead{Model} &
        \colhead{$T$}   &
        \colhead{$\theta_0$}   &
        \colhead{$\Gamma_0$}   &
        \colhead{Luminosity ($L_j$)}   &
        \colhead{Inner Boundary Radius}   &
        \colhead{$\Gamma_{\infty}$}   &
        \colhead{$t_{\rm bo}$}     \\
        \colhead{}   &   \colhead{(s)}   &  \colhead{}   &  \colhead{}   &  \colhead{(erg s$^{-1}$)}   &  \colhead{($r_0$)}   &  \colhead{}   &        		\colhead{(s)}   
		}
\startdata
m5Tcon &   0.0$^{a}$		&		&		&		&		&		& 4.2 \\
m5T0.2 &		0.2		&		&		&		&		&		& 4.86 \\
m5T1.0 &		1.0		&	5$\degree$	&	5.0	&	$5 \times 10^{50}$ &	$4 \times 10^8$ cm	&	400	& 5.8 \\
m5T2.0 &		2.0		&		&		&		&		&		& 6.1 \\
m5T4.0 &		4.0		&		&		&		&		&		& 4.45 \\ \hline
m10Tcon &   0.0		&		&		&		&		&		& 4.4 \\
m10T0.2 &		0.2		&		&		&		&		&		& 6.47 \\
m10T1.0 &		1.0		&	10$\degree$	&	5.0	&	$5 \times 10^{50}$ &	$4 \times 10^8$ cm	&	400	& 6.0 \\
m10T2.0 &		2.0		&		&		&		&		&		& 6.16 \\
m10T4.0 &		4.0		&		&		&		&		&		& 4.88
\enddata
\tablenotetext{a}{$T=0.0$ means the jet is continuous.}
\end{deluxetable}

\begin{figure}
\centering
   \subfloat{\includegraphics[width=0.6\linewidth]{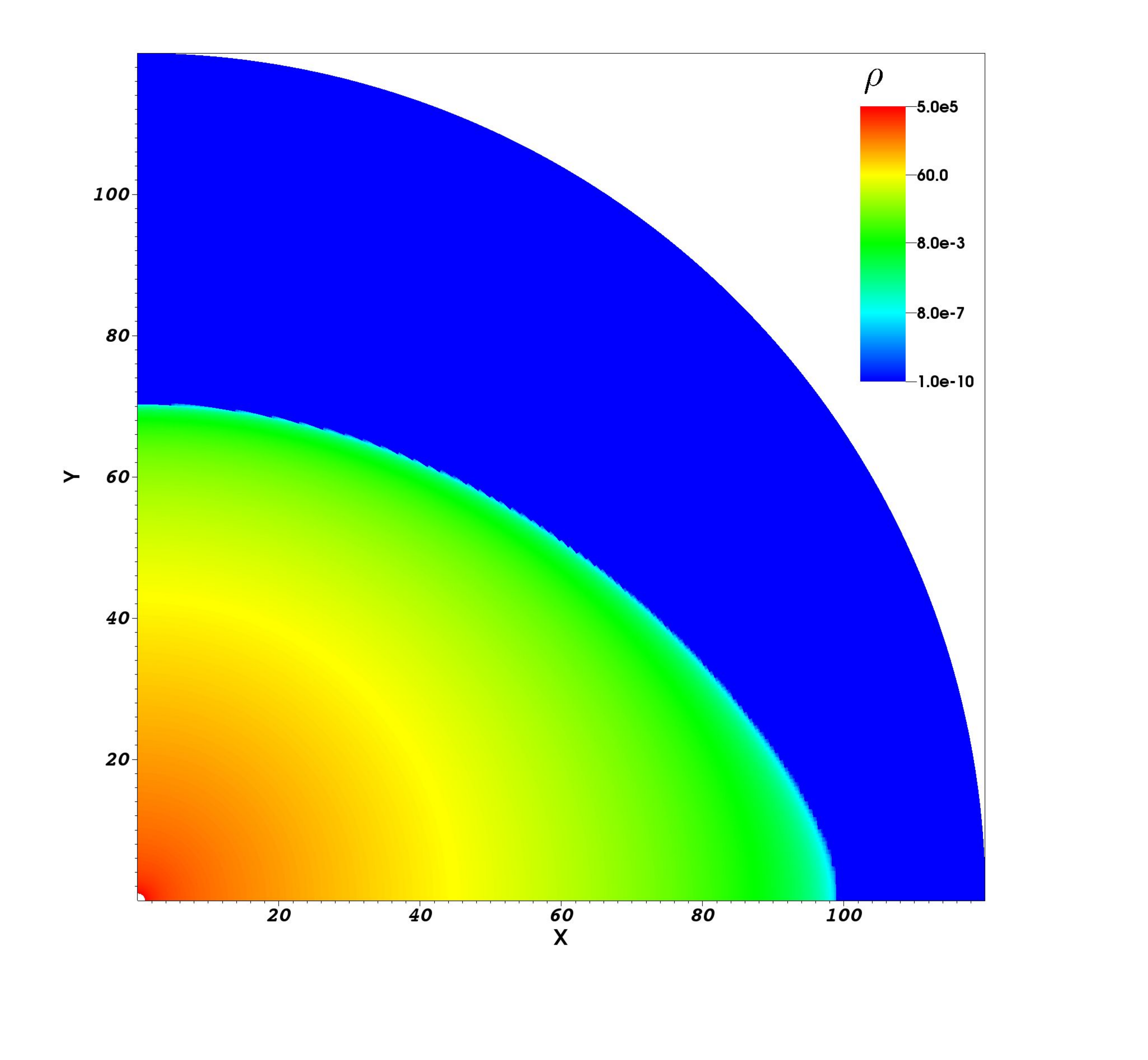}}\\
   \subfloat{\includegraphics[width=0.6\linewidth]{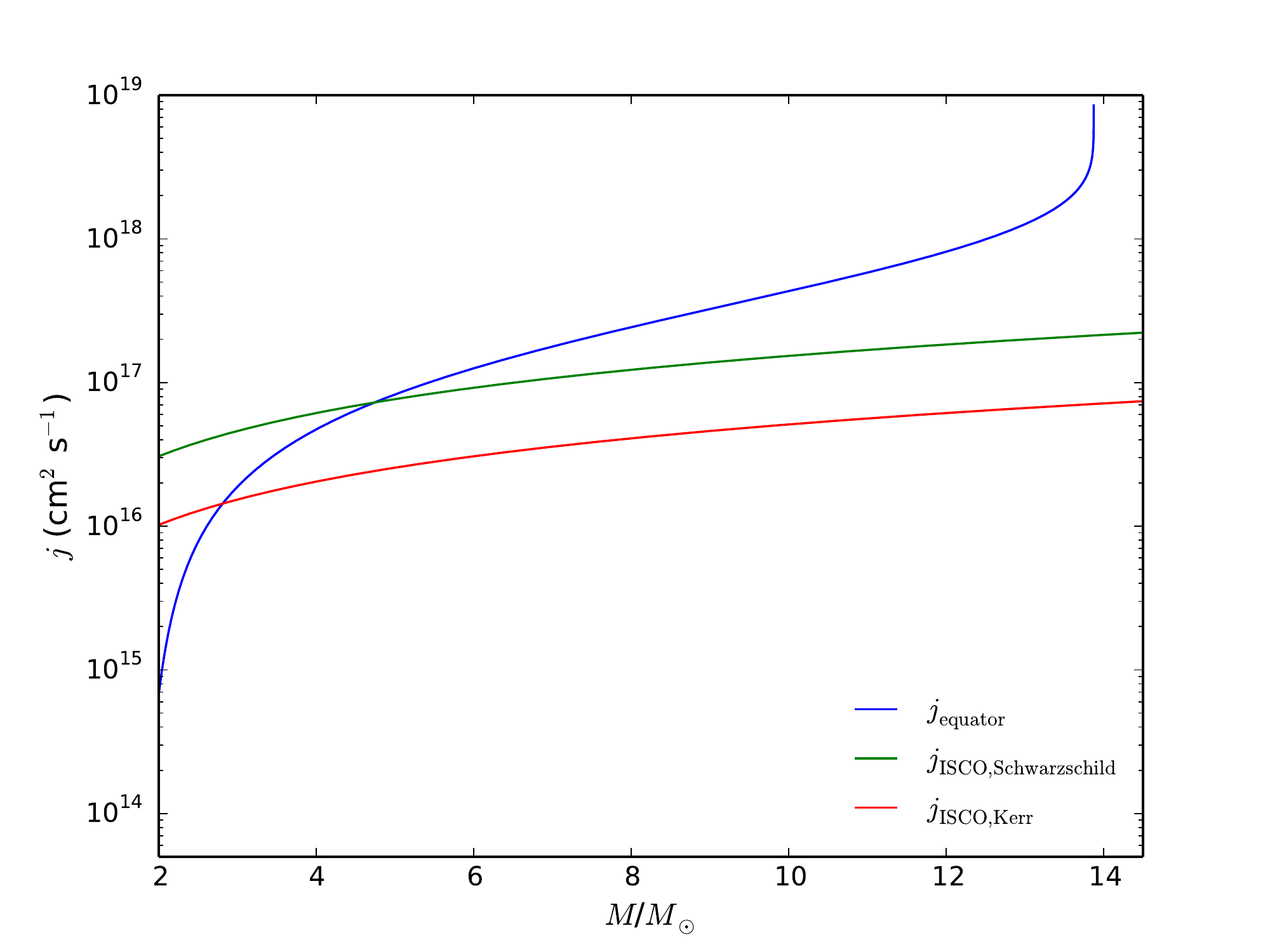}}
    \caption{Upper panel: the density contour in the meridian section of the star envelope.
   Density $\rho$ is in units of g~cm$^{-3}$, the X and Y (note Y is referred to as $z$ in the paper) axis unit scale is $4 \times 10^8$~cm.
   Lower panel: the distribution of the equatorial specific angular momentum along the star mass in our paper (blue solid line).
   The green line indicates the specific angular momentum at the ISCO for a Schwarzschild black hole 
   as a function of their masses. The red line shows the specific angular momentum at the ISCO for a Kerr black hole
   (with rotational parameter $a = 1$) as a function of their masses.}
    \label{Fig:plot1}
\end{figure}

\begin{figure}
\centering
   \subfloat{\includegraphics[width=0.3\linewidth]{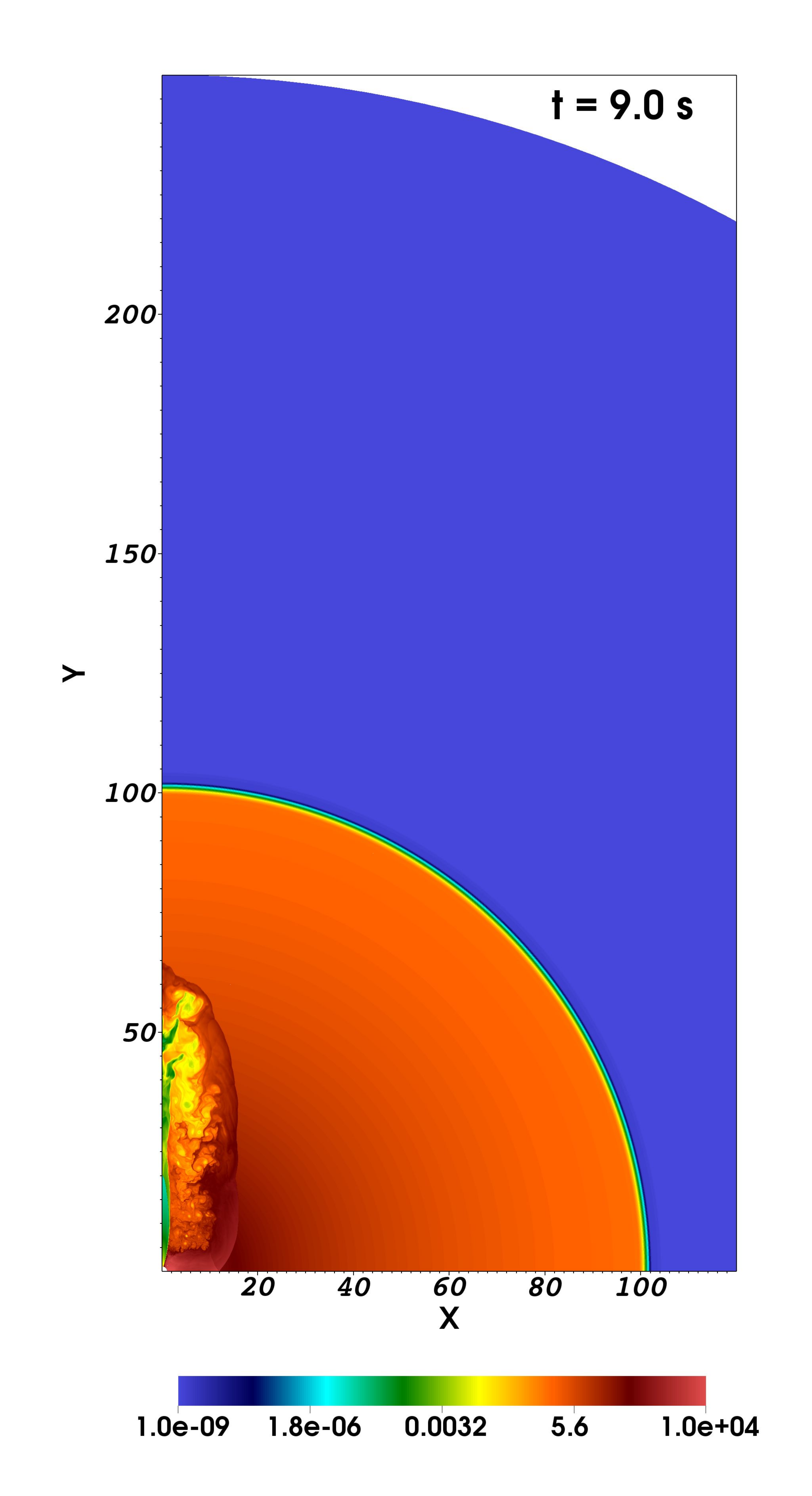}}
   \subfloat{\includegraphics[width=0.3\linewidth]{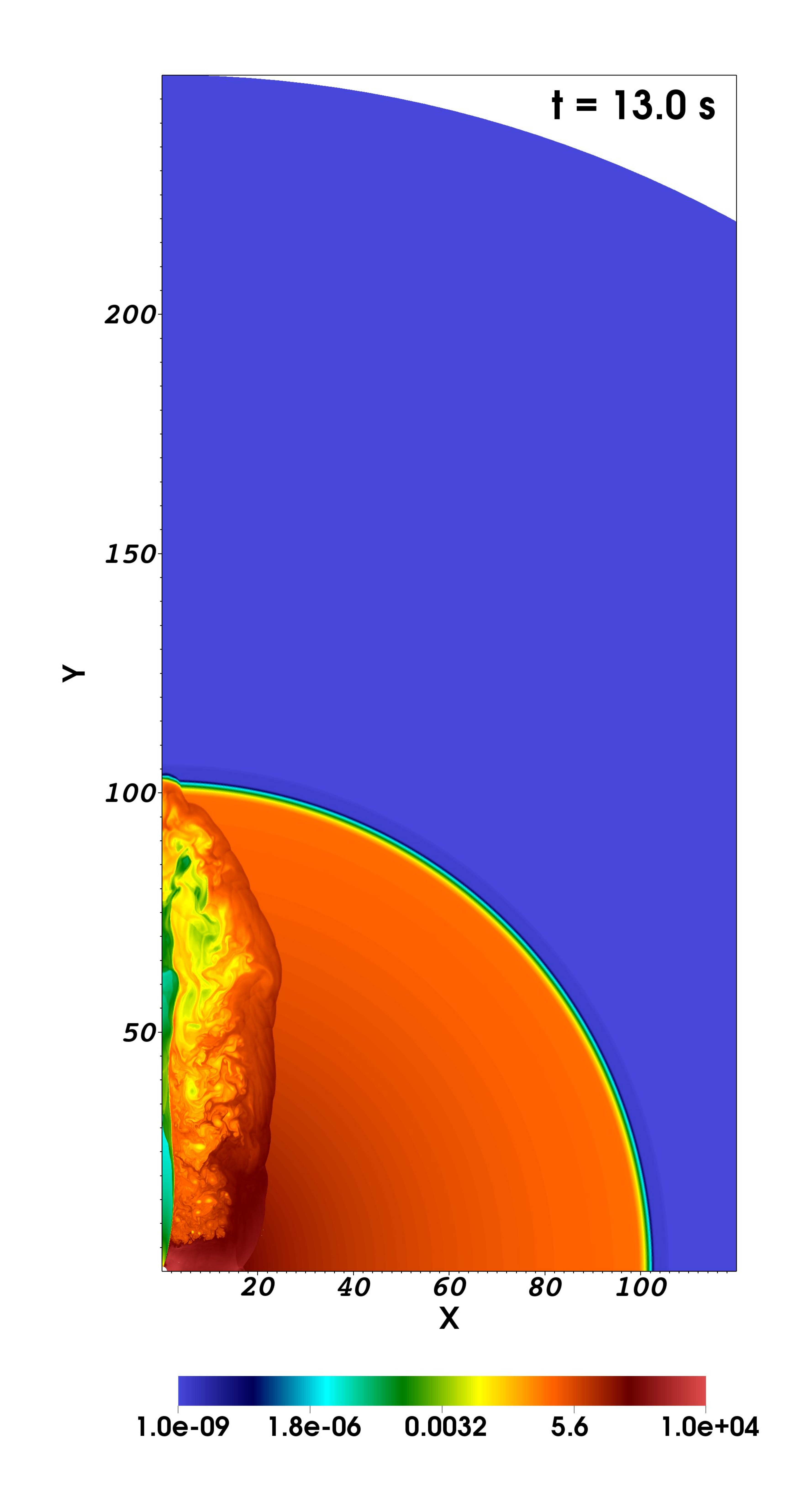}}
   \subfloat{\includegraphics[width=0.3\linewidth]{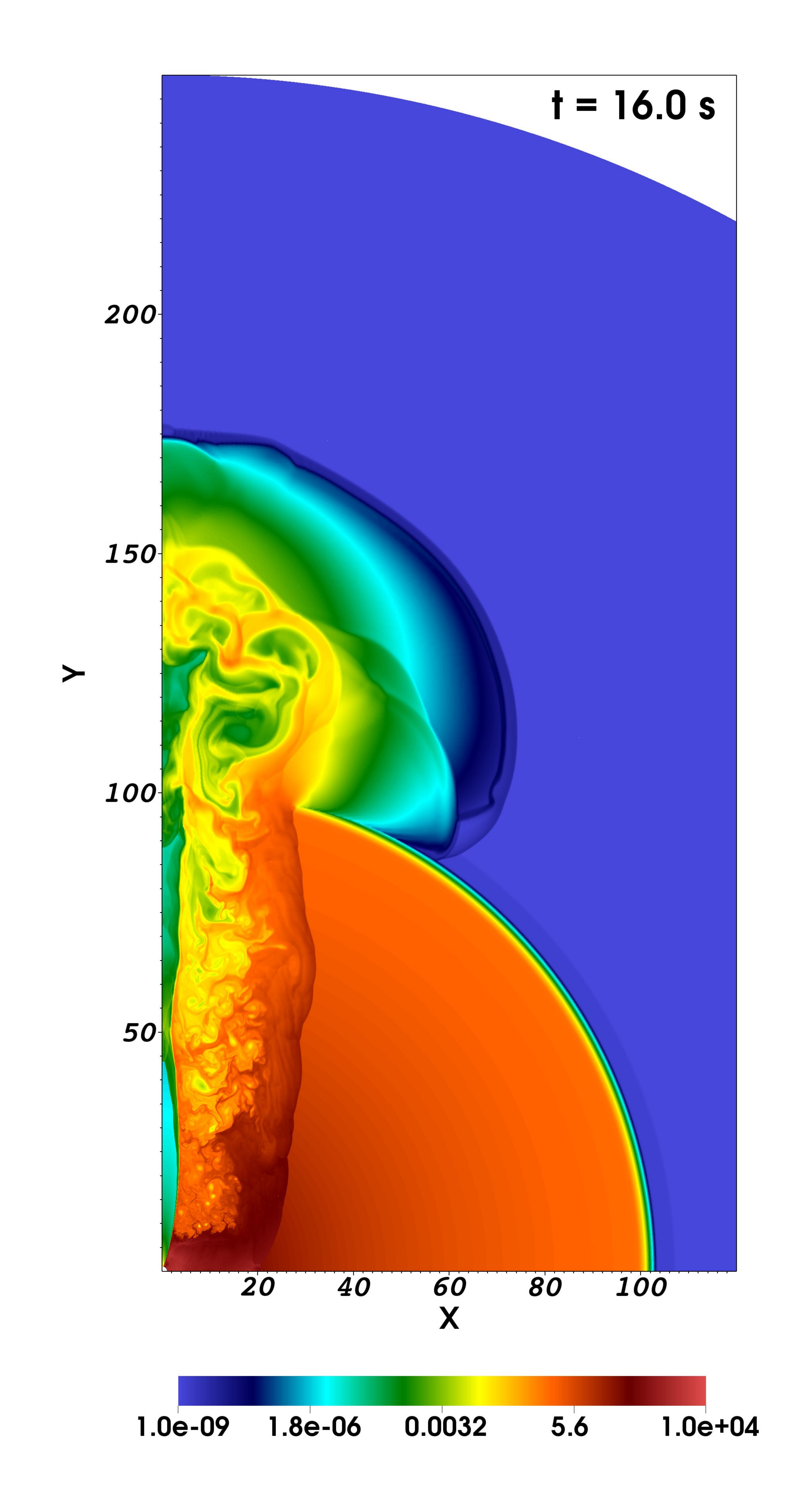}} \\
   \subfloat{\includegraphics[width=0.3\linewidth]{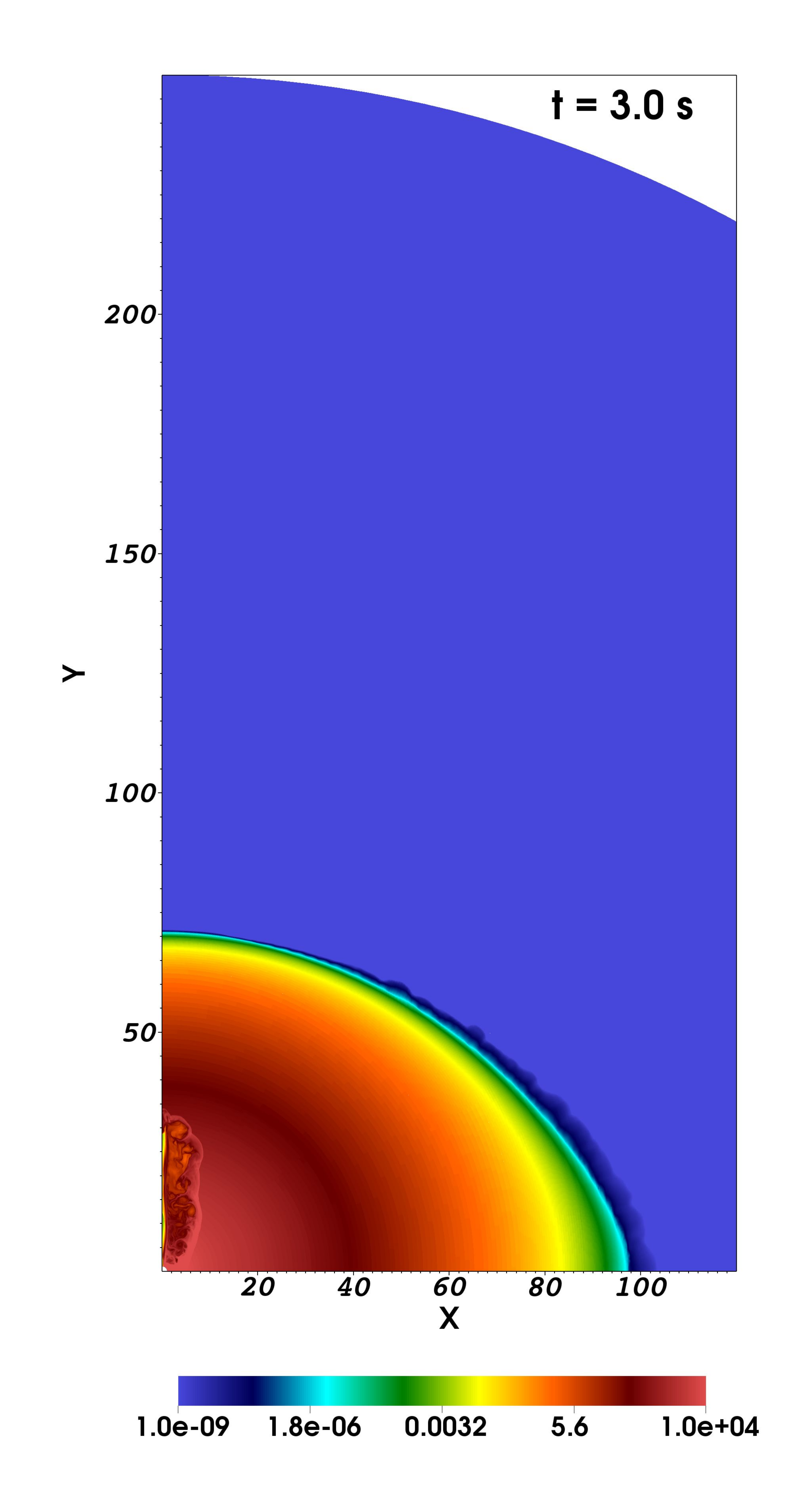}}
   \subfloat{\includegraphics[width=0.3\linewidth]{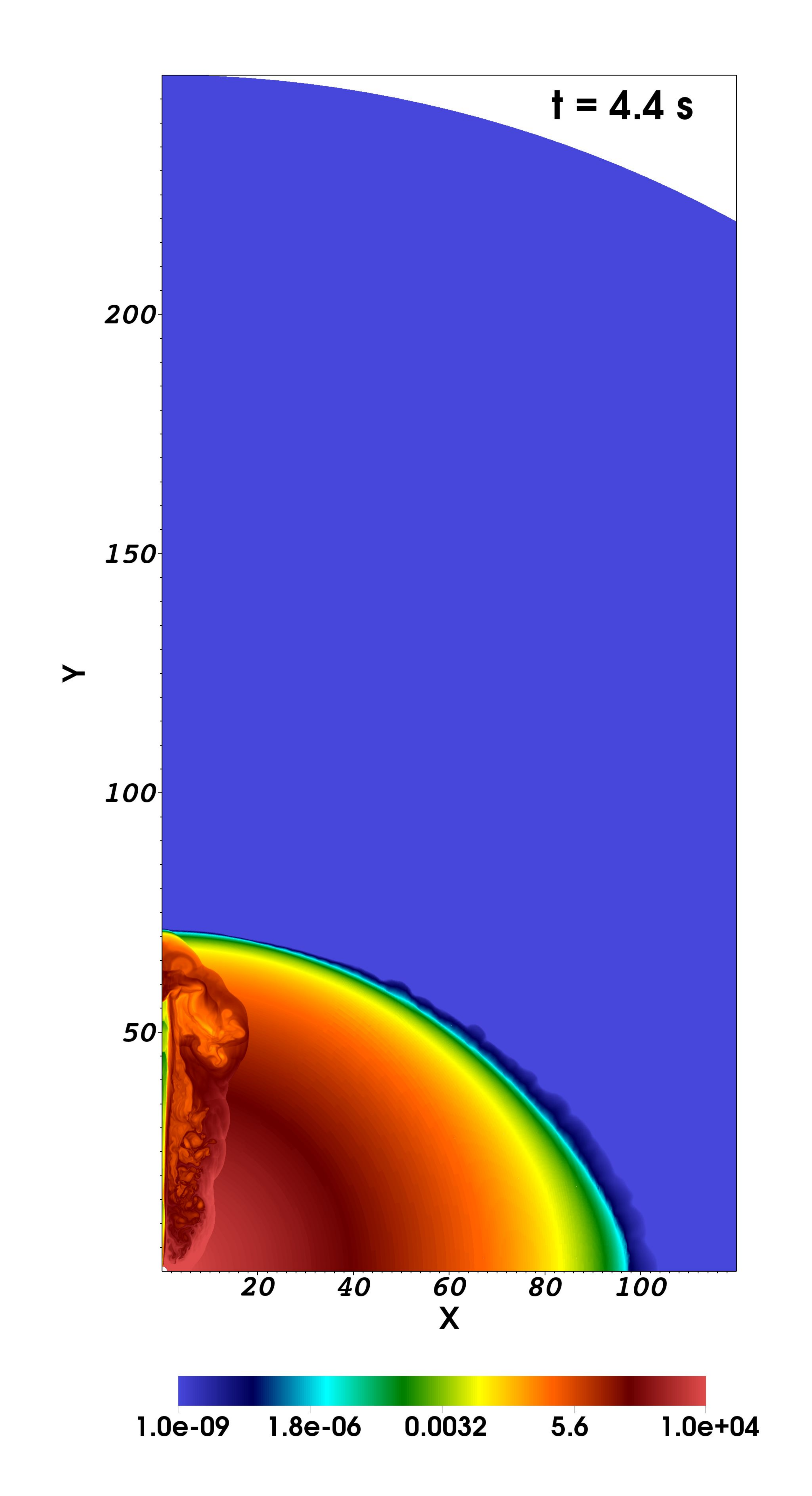}}
   \subfloat{\includegraphics[width=0.3\linewidth]{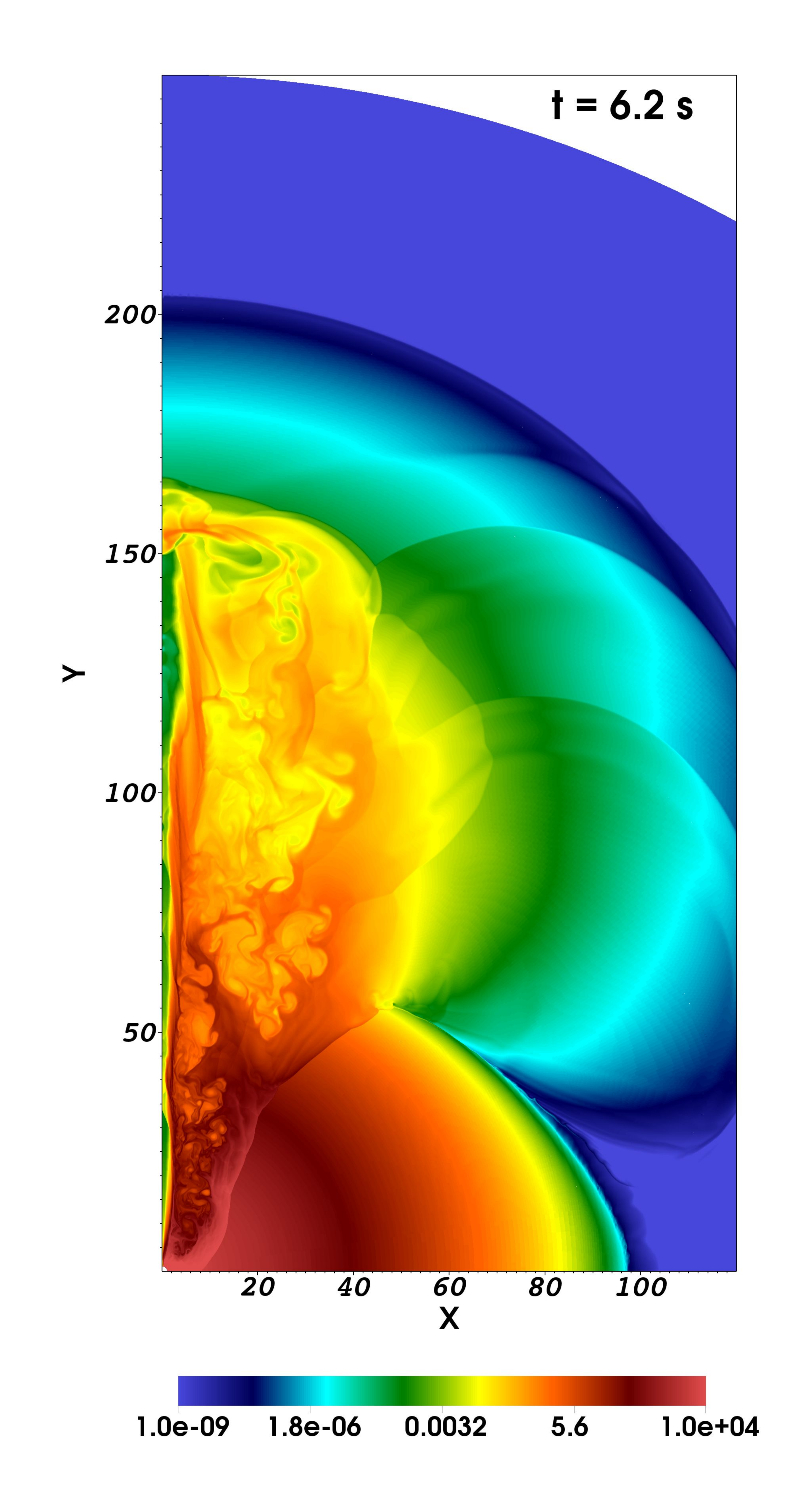}}
\caption{Time sequence of density distribution for the propagation of a continuous jet in a star. 
Upper panel: the case of spherical non-rotating star,
the unit scale for X and Y axis is $10^9$~cm.
The physical initial conditions are the same to the model t10g5 in \cite{Morsony07}.
Lower panel: the case of a rotating star in the model m10Tcon,
the unit scale for X and Y axis is $4 \times 10^8$~cm.
Density is in units of g~cm$^{-3}$.}
\label{Fig:plot2}
\end{figure}

\begin{figure}
   \begin{center}
   \includegraphics[scale=0.4]{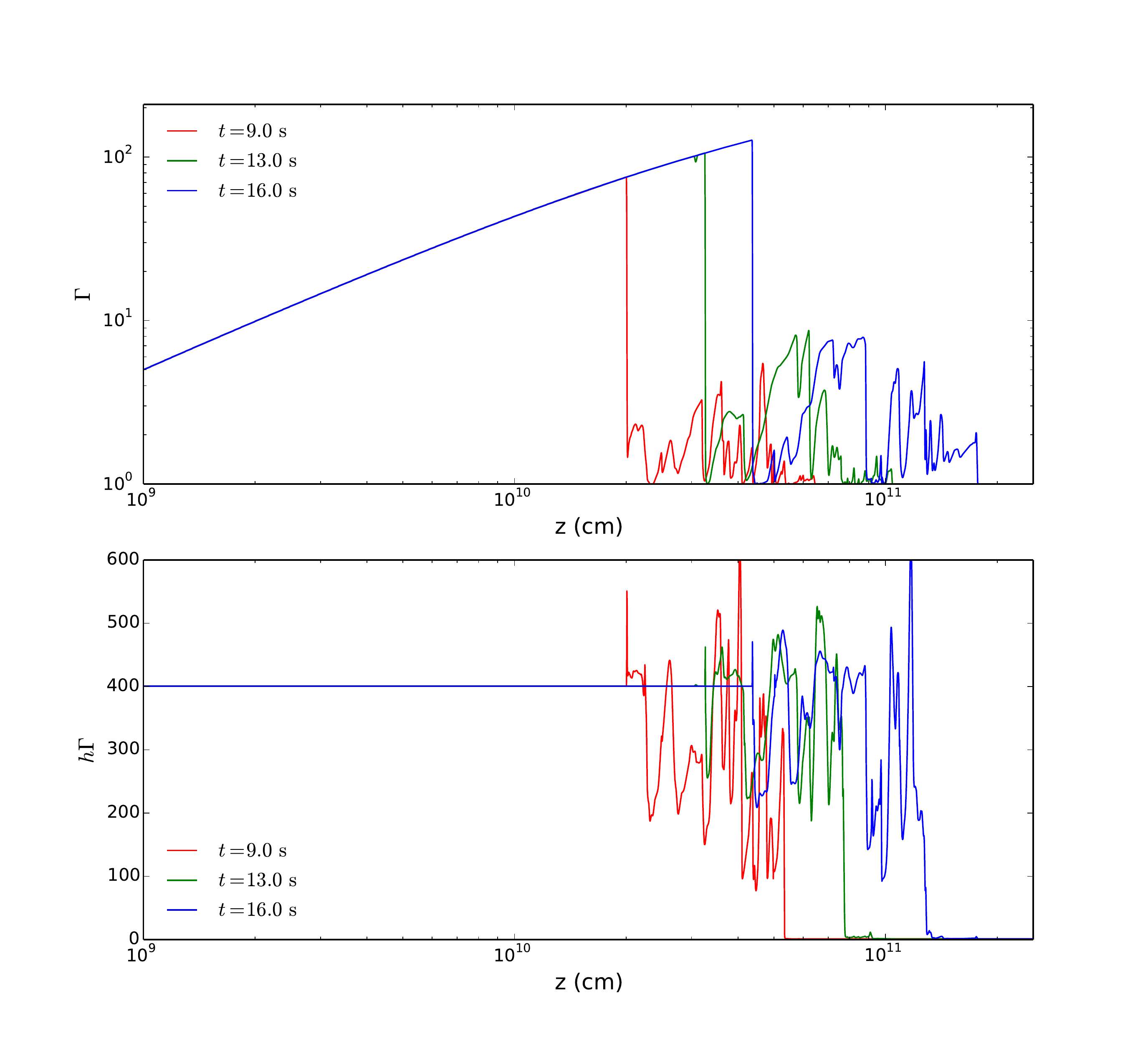}
   \caption{Lorentz factor (upper panel) and Bernoulli's constant $h \Gamma$ (lower panel) along the jet axis
   at $t =$ 9.0, 13.0, and 16.0 s respectively.
   }
   \label{Fig:plot3}
   \end{center}
\end{figure}

\begin{figure}
   \begin{center}
   \includegraphics[scale=0.5]{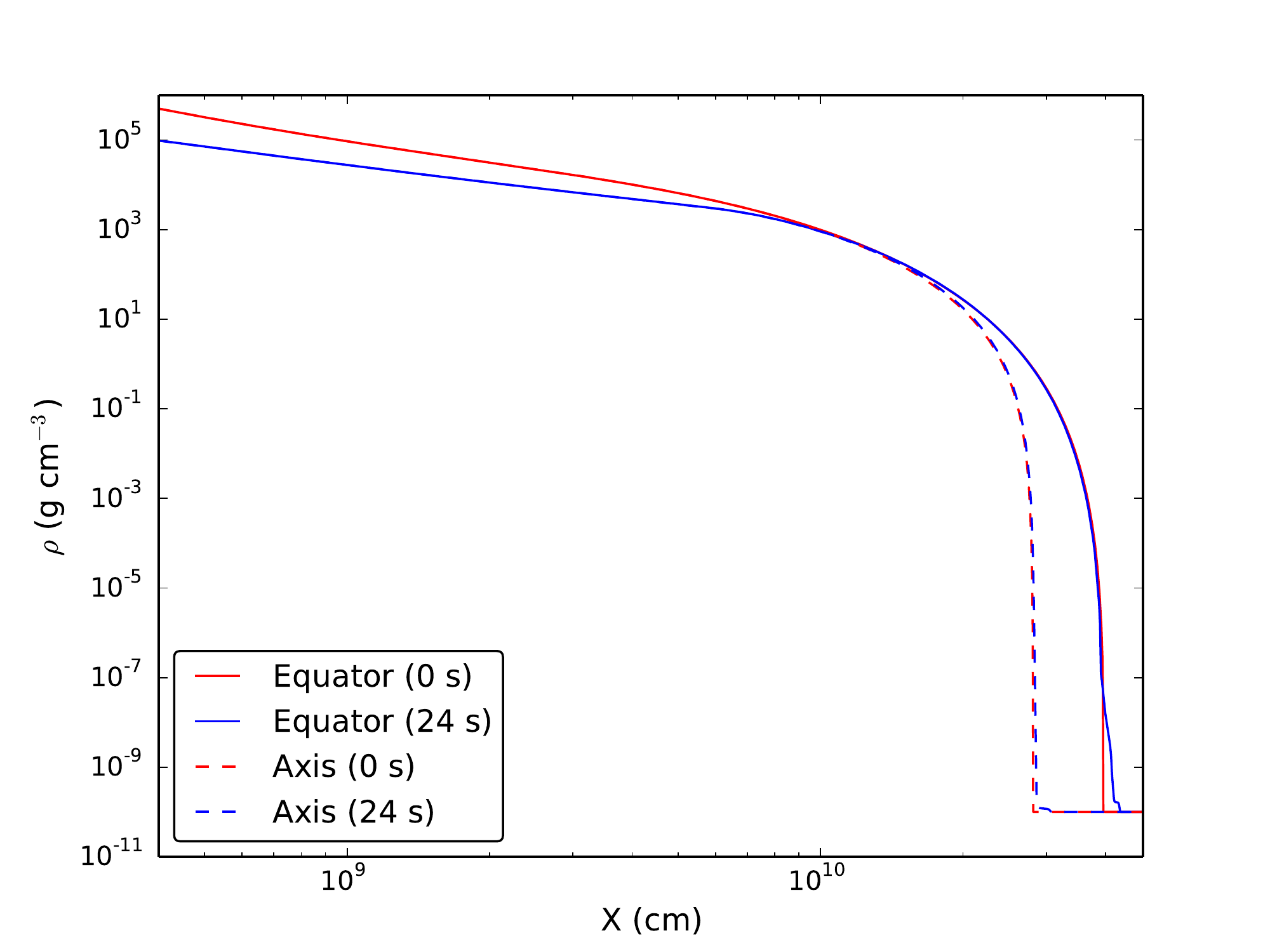}
   \caption{Evolution of the density profiles in the equatorial (solid lines) and axial (dashed lines) directions at $t=0$ s
   (red) and 24 s (blue) after the collapse starts.
   }
   \label{Fig:plot4}
   \end{center}
\end{figure}

\begin{figure}
\centering
   \subfloat{\includegraphics[width=0.4\linewidth]{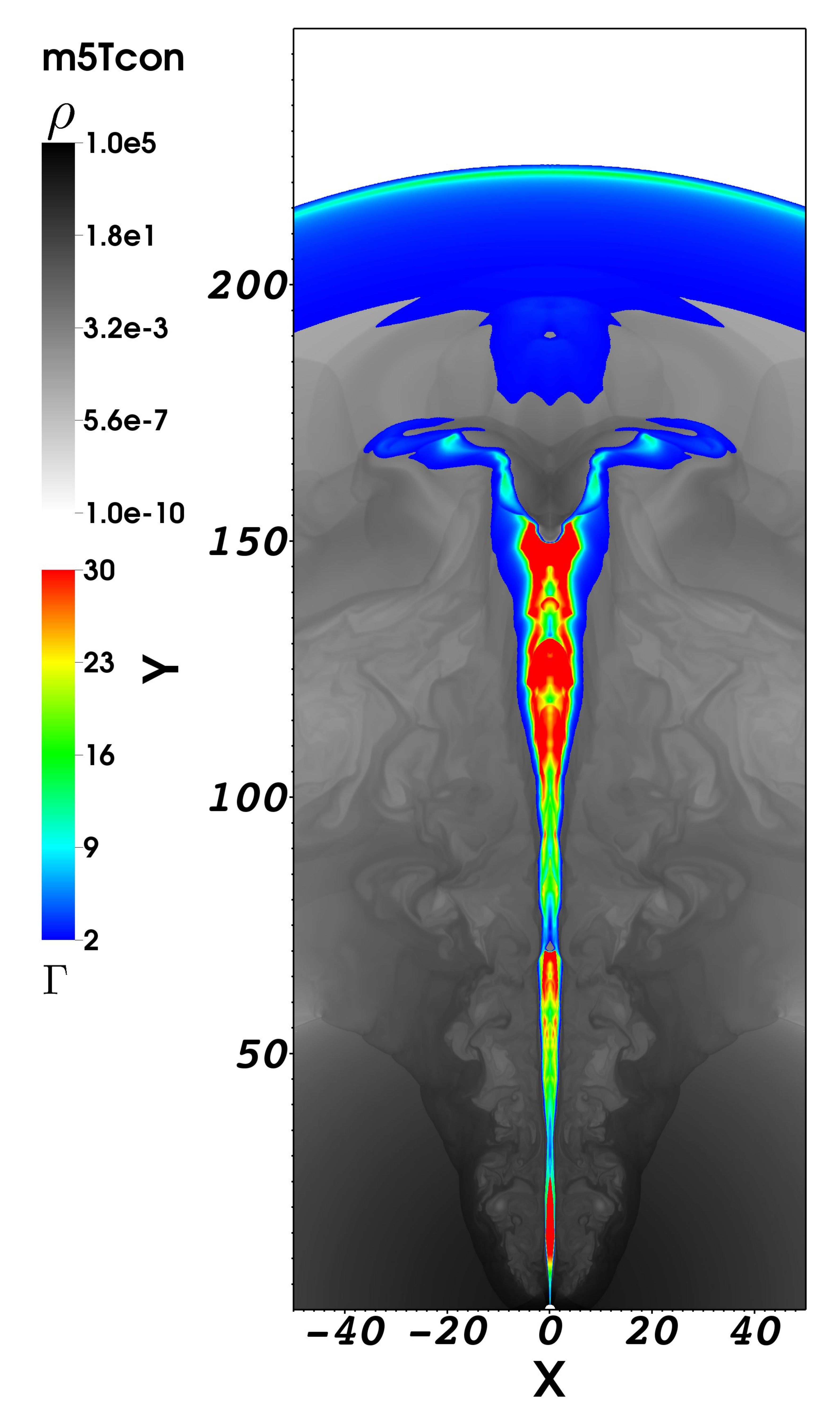}}
   \subfloat{\includegraphics[width=0.4\linewidth]{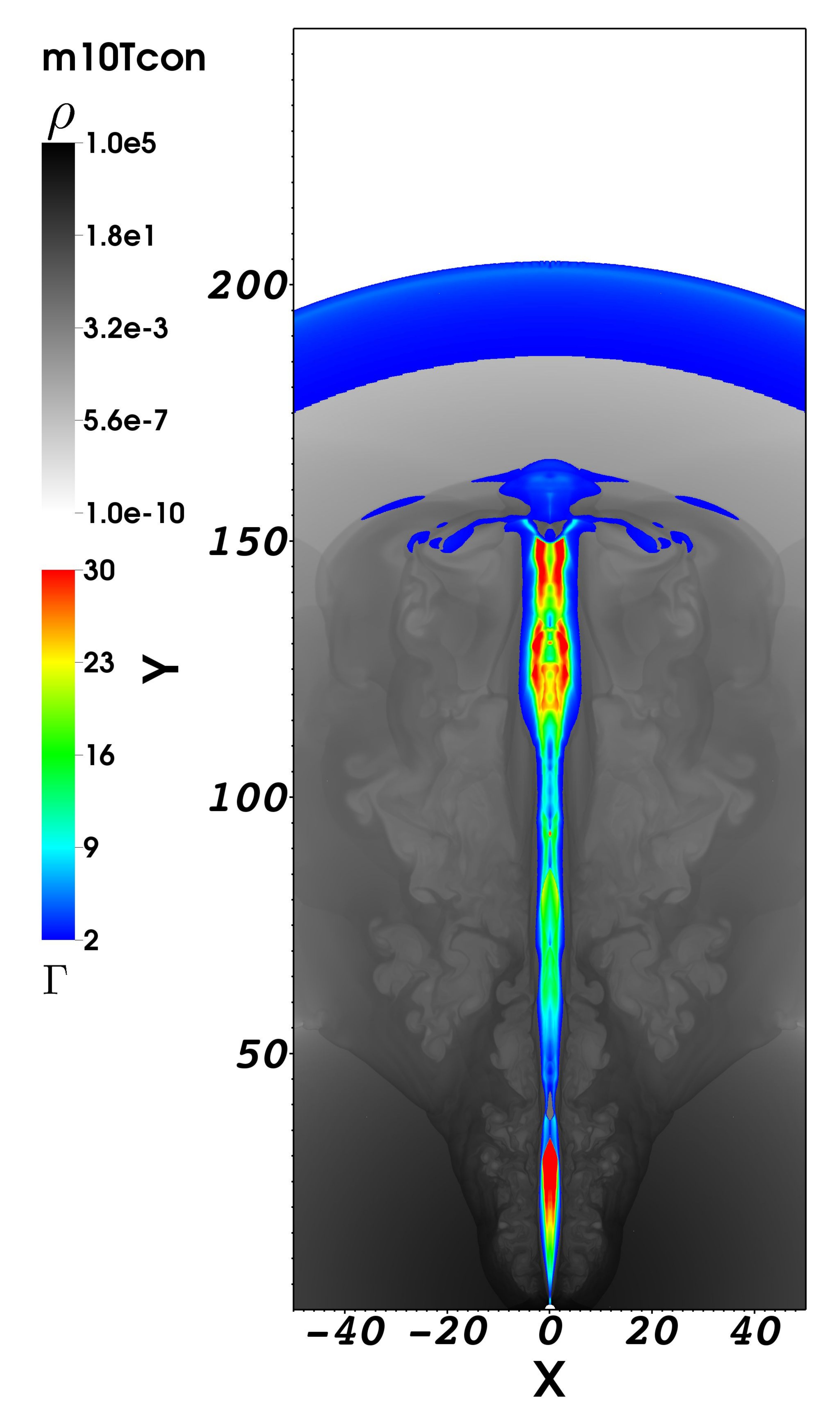}}
\caption{Lorentz factor map and density distribution (in units of g~cm$^{-3}$)
of model m5Tcon at $t = 6.2$ s (left panel), and model m10Tcon at $t = 6.2$ s (right panel).
The unit scale for X and Y axis is $4 \times 10^8$~cm.}
\label{Fig:plot5}
\end{figure}

\begin{figure}
   \begin{center}
   \includegraphics[scale=0.5]{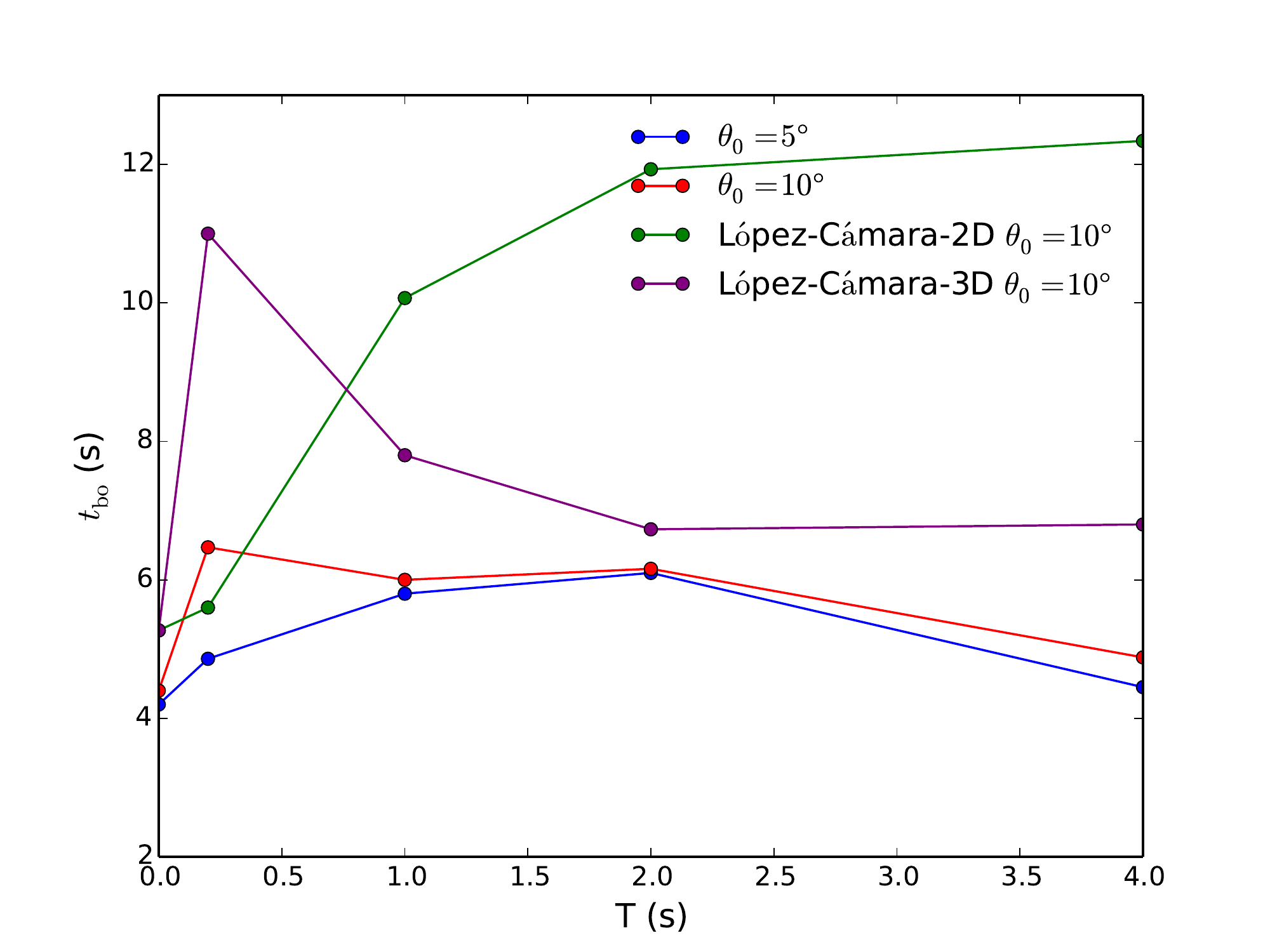}
   \caption{The $t_{\rm bo}-T$ relation for two series of simulations
   (blue dots for $\theta_0 = 5 \degree$, red dots for $\theta_0 = 10 \degree$) in this work.
   For comparison, the green and purple dots represent the 2D and 3D results from 
   \cite{Lopez16}, respectively.
   }
   \label{Fig:plot6}
   \end{center}
\end{figure}

\begin{figure}
\centering
   \subfloat{\includegraphics[width=0.35\linewidth]{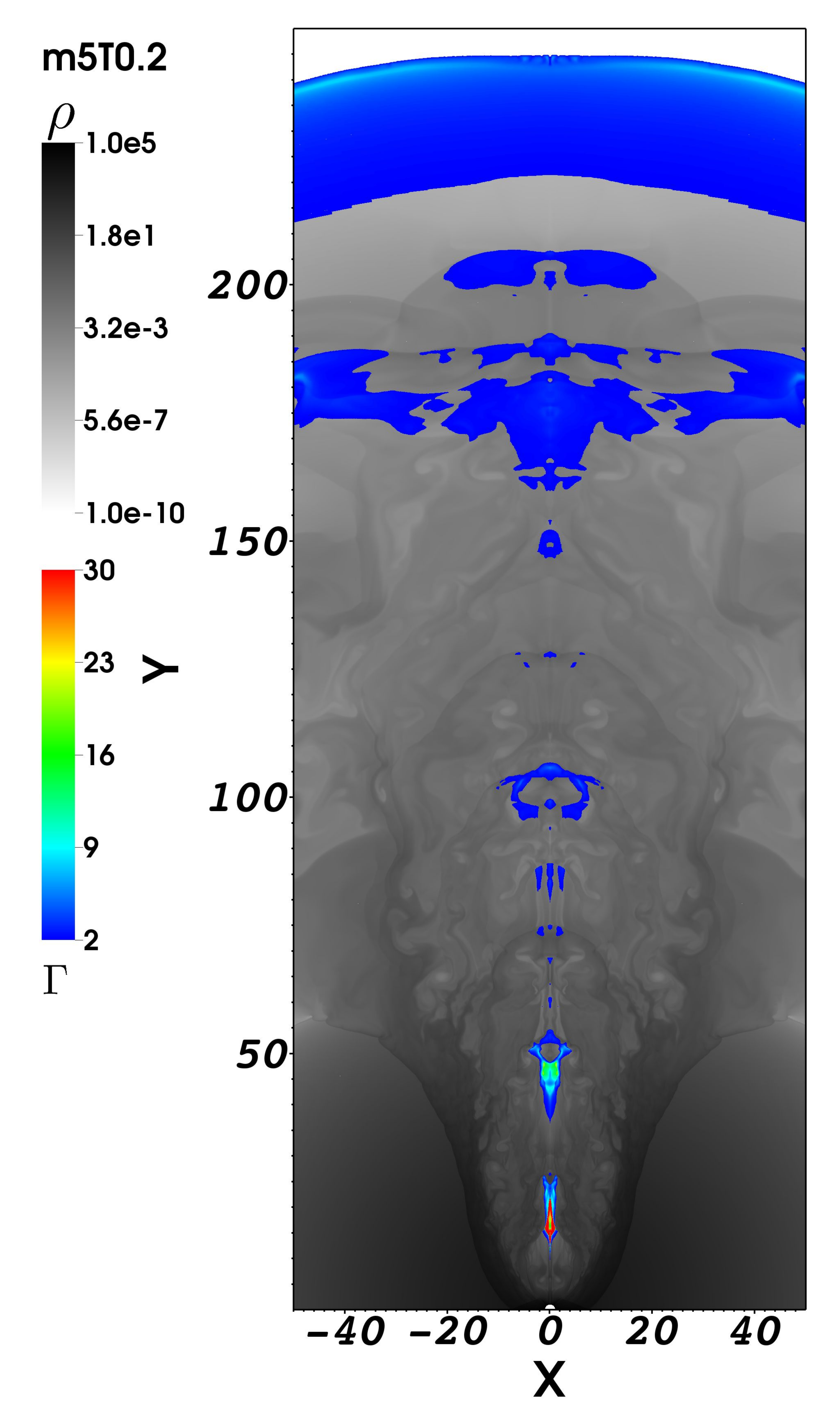}}
   \subfloat{\includegraphics[width=0.35\linewidth]{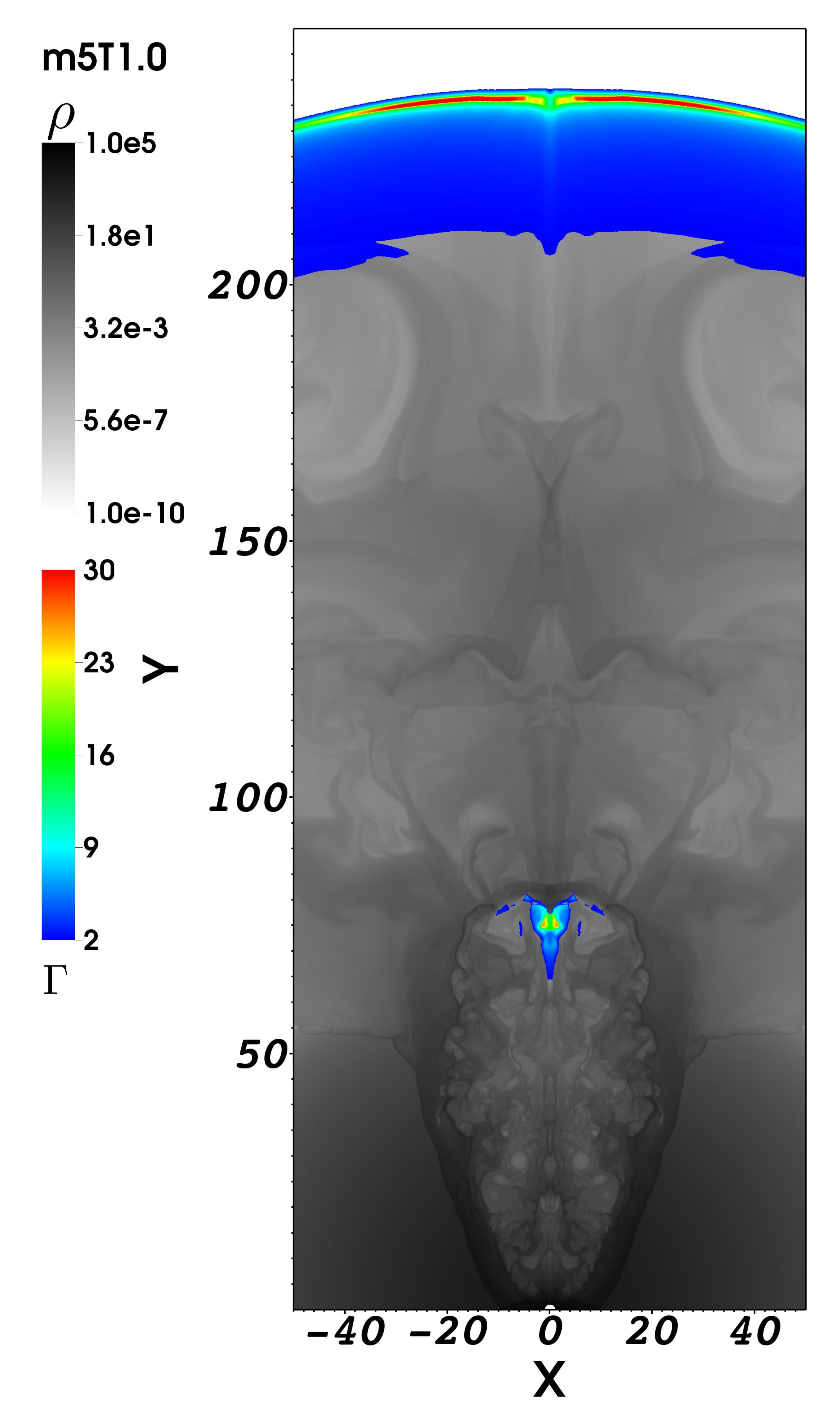}} \\
   \subfloat{\includegraphics[width=0.35\linewidth]{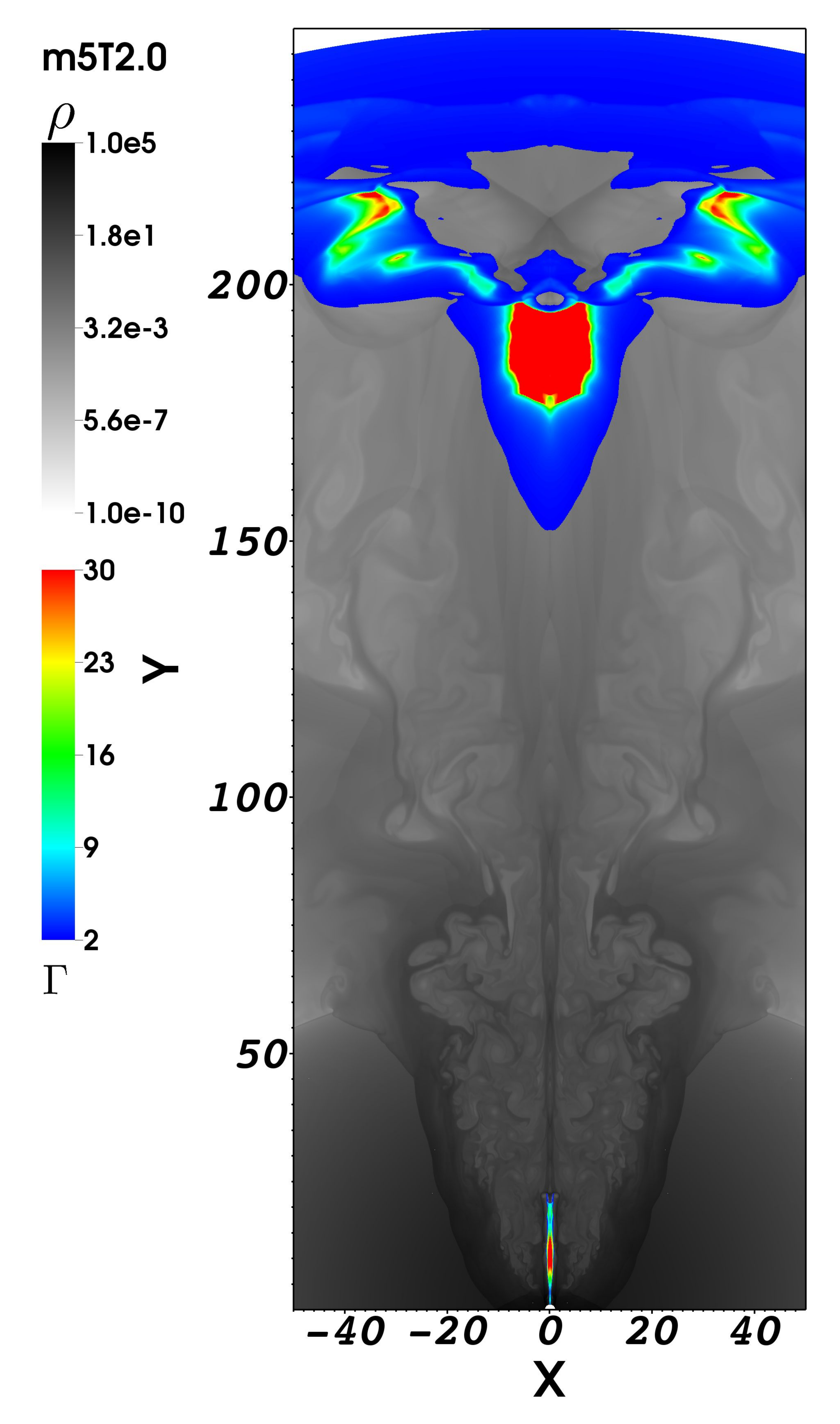}}
   \subfloat{\includegraphics[width=0.35\linewidth]{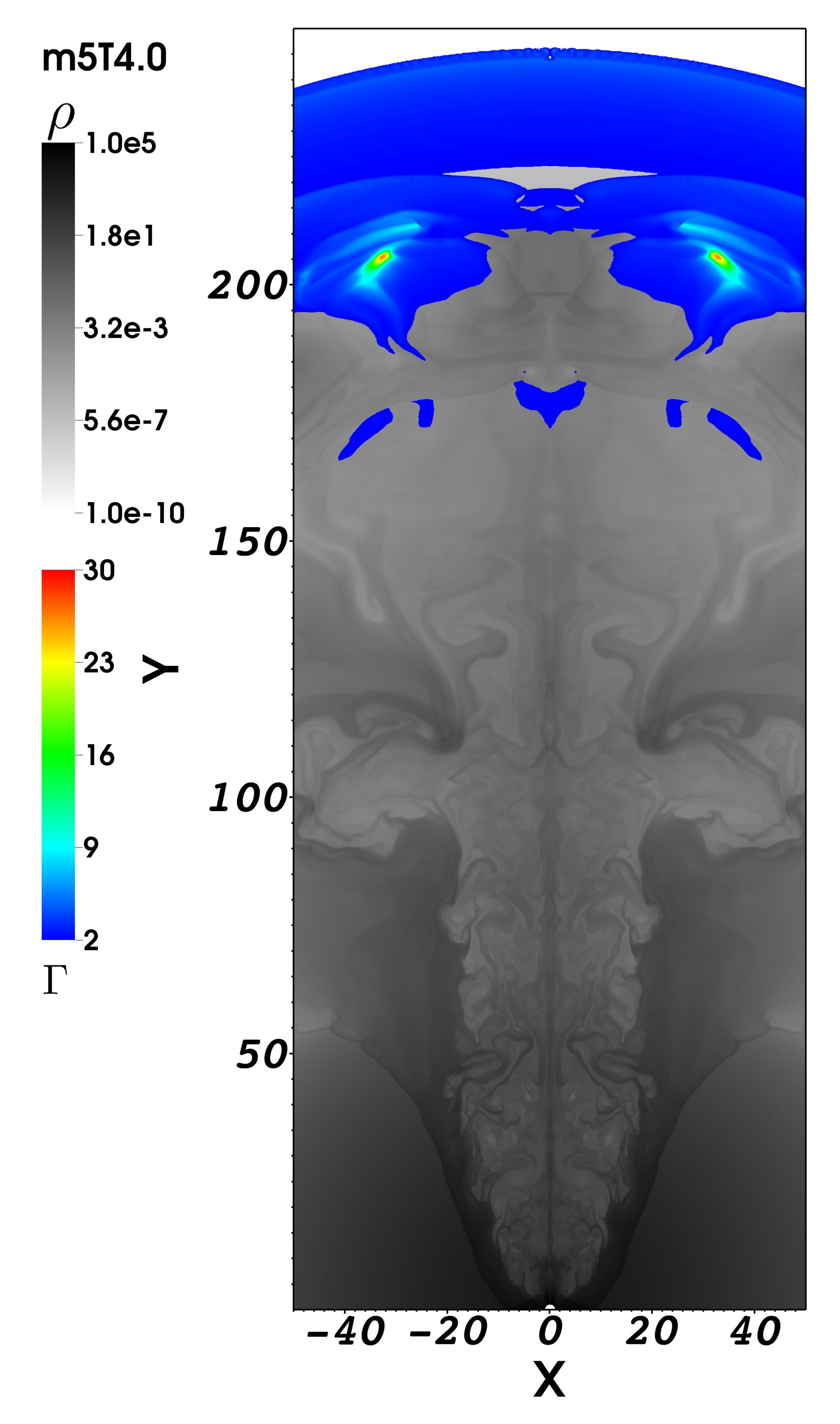}}
\caption{Lorentz factor map and density distribution (in units g~cm$^{-3}$)
of model m5T0.2 at $t = 7.2$ s (upper-left panel), model m5T1.0 at $t = 8.0$ s (upper-right panel), 
model m5T2.0 at $t = 8.4$ s (lower-left panel), and model m5T4.0 at $t = 6.8$ s (lower-right panel).
The unit scale for X and Y axis is $4 \times 10^8$~cm.}
\label{Fig:plot7}
\end{figure}

\begin{figure}
\centering
   \subfloat{\includegraphics[width=0.35\linewidth]{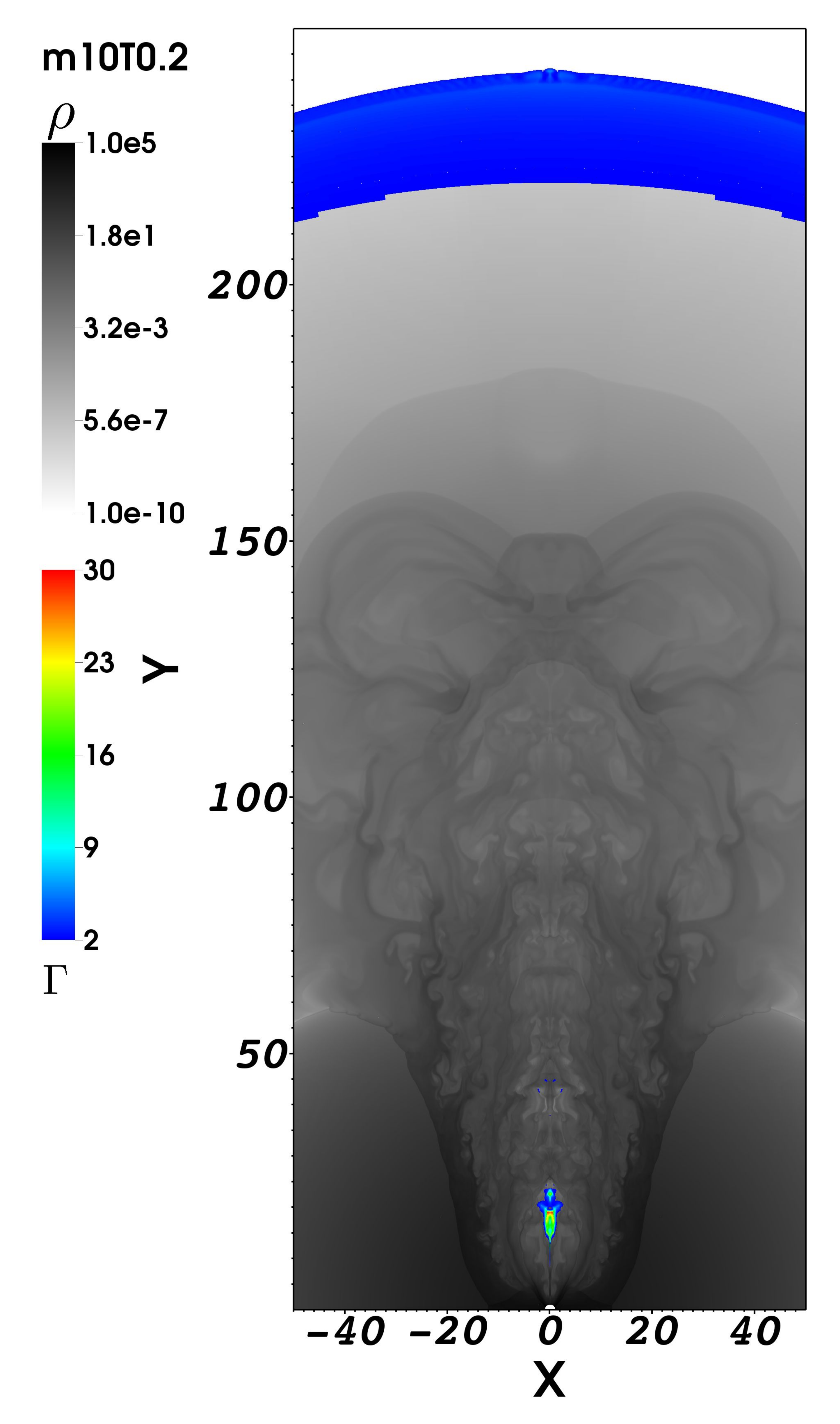}}
   \subfloat{\includegraphics[width=0.35\linewidth]{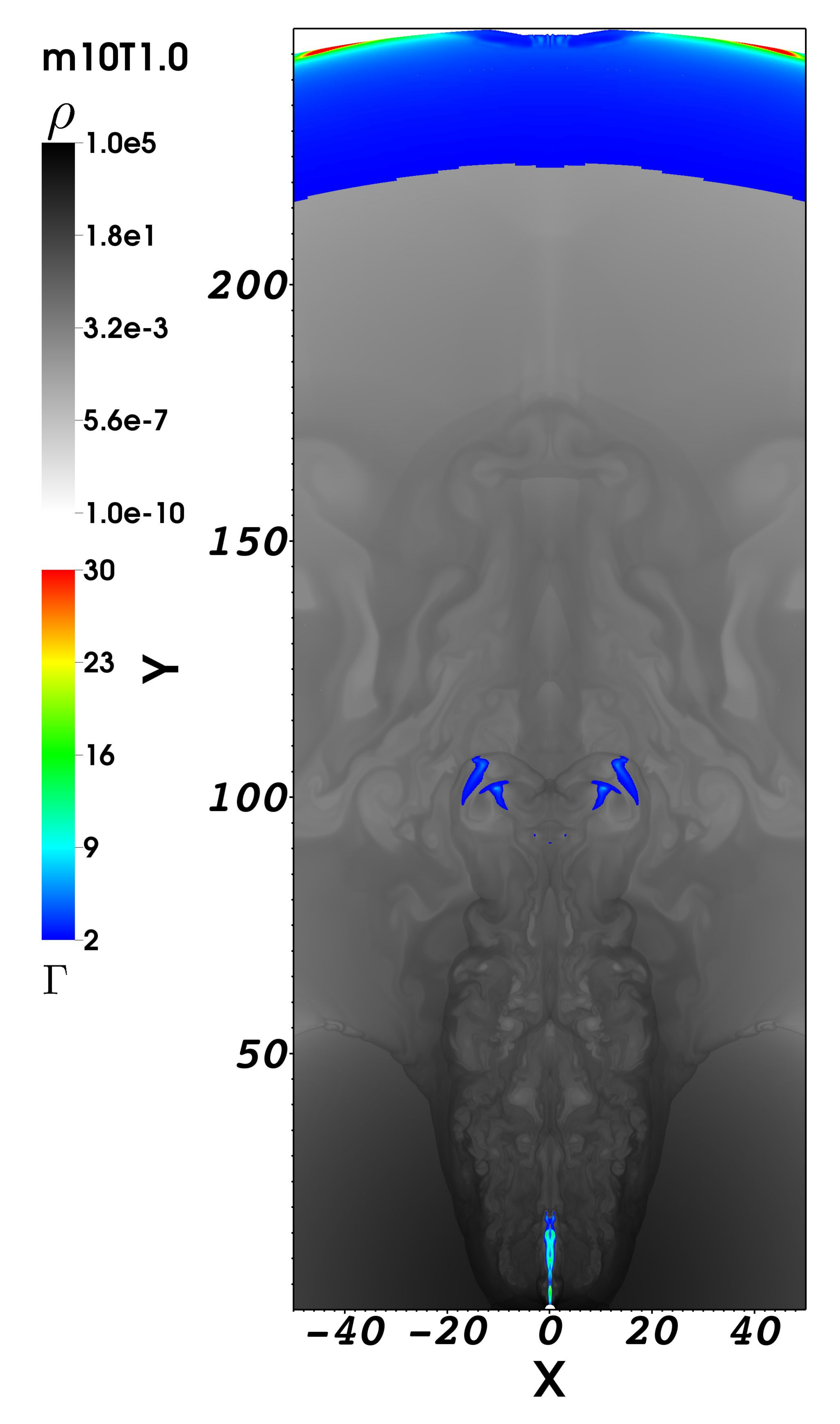}} \\
   \subfloat{\includegraphics[width=0.35\linewidth]{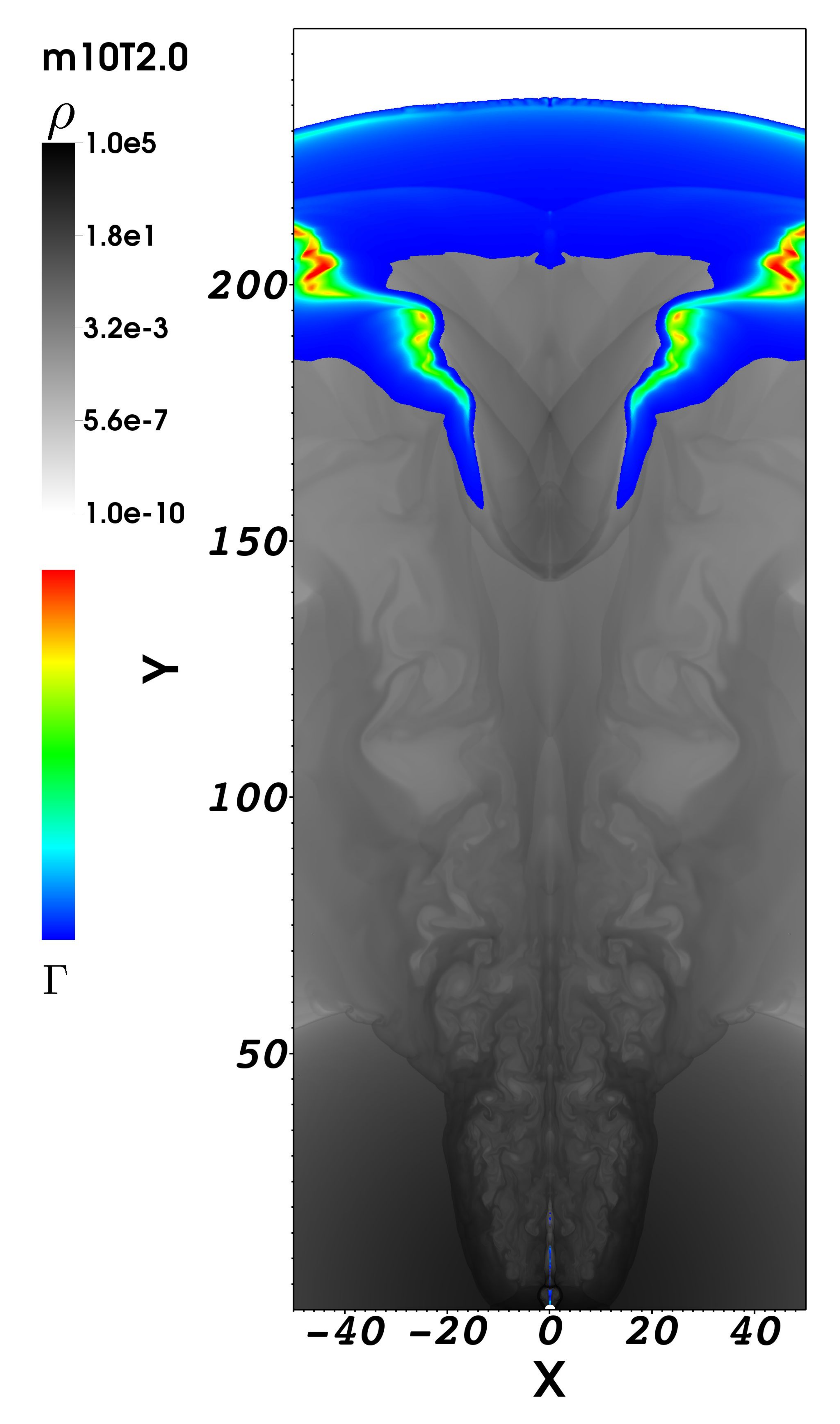}}
   \subfloat{\includegraphics[width=0.35\linewidth]{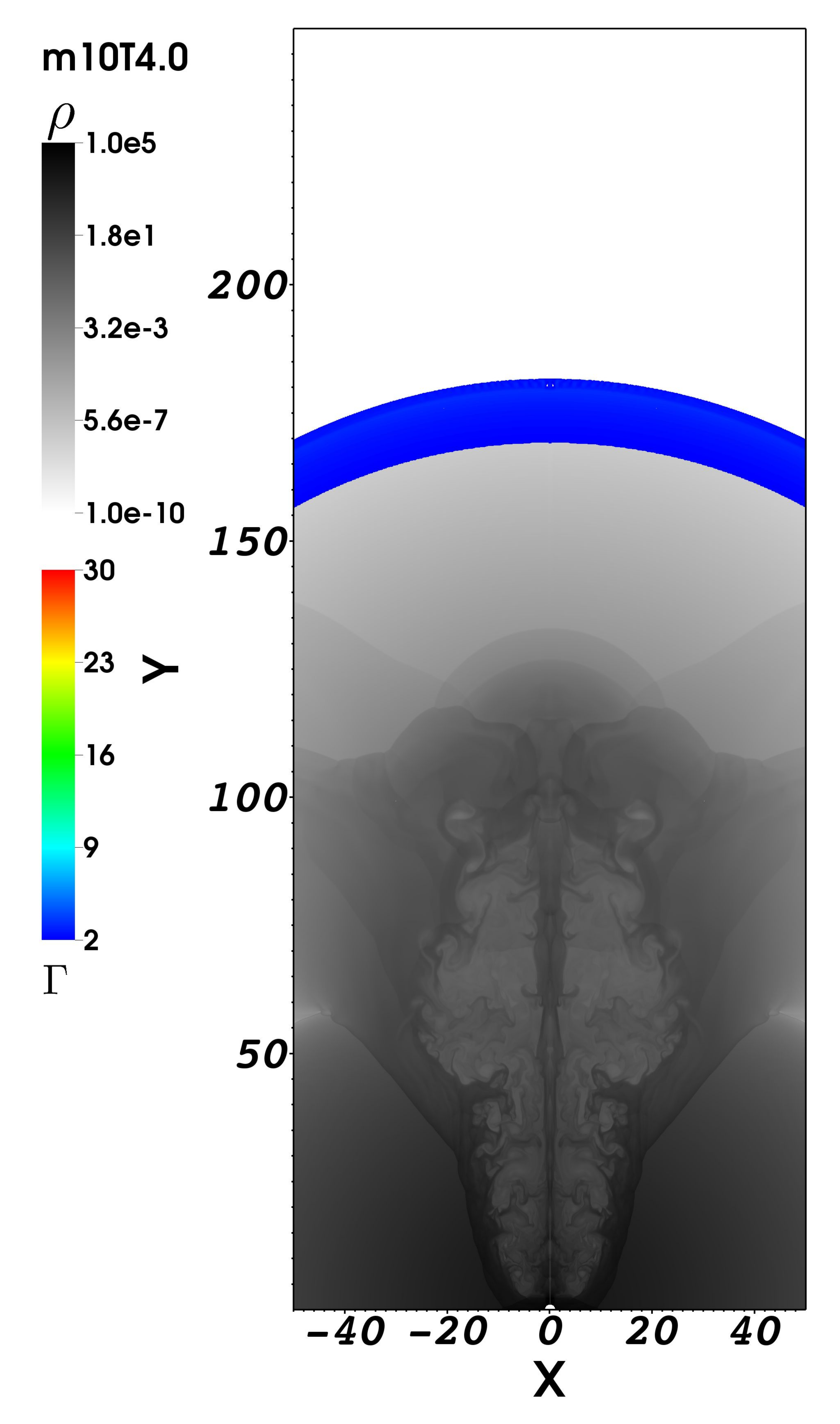}}
\caption{Lorentz factor map and density distribution (in units of g~cm$^{-3}$)
of model m10T0.2 at $t = 8.8$ s (upper-left panel), model m10T1.0 at $t = 8.4$ s (upper-right panel), 
model m10T2.0 at $t = 8.4$ s (lower-left panel), and model m10T4.0 at $t = 6.4$ s (lower-right panel).
The unit scale for X and Y axis is $4 \times 10^8$~cm.}
\label{Fig:plot8}
\end{figure}

\begin{figure}
\centering 
 \subfloat{\includegraphics[width=0.9\linewidth]{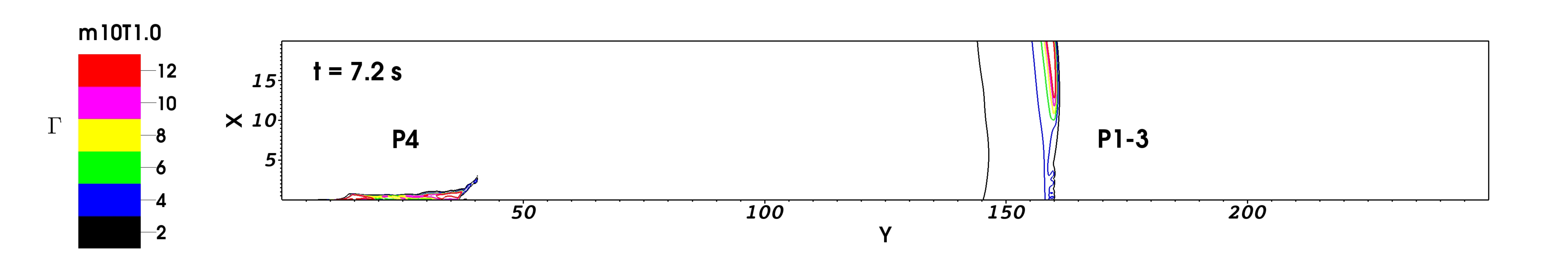}} \\
 \subfloat{\includegraphics[width=0.9\linewidth]{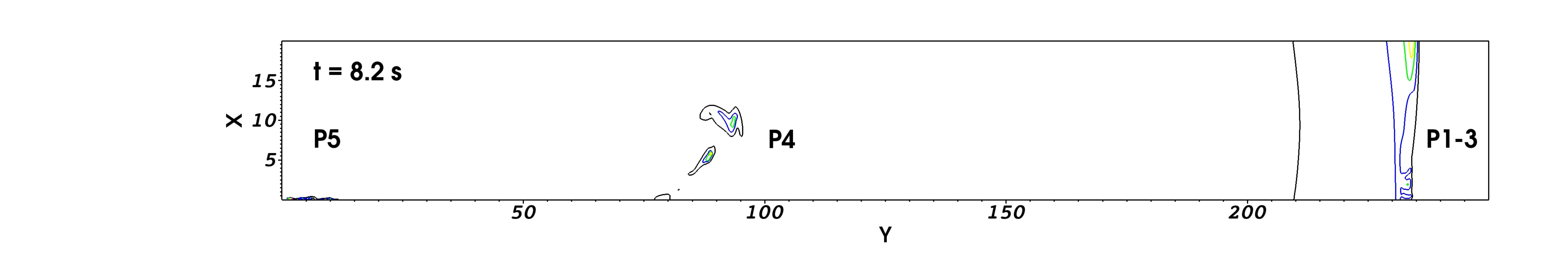}} \\
 \subfloat{\includegraphics[width=0.9\linewidth]{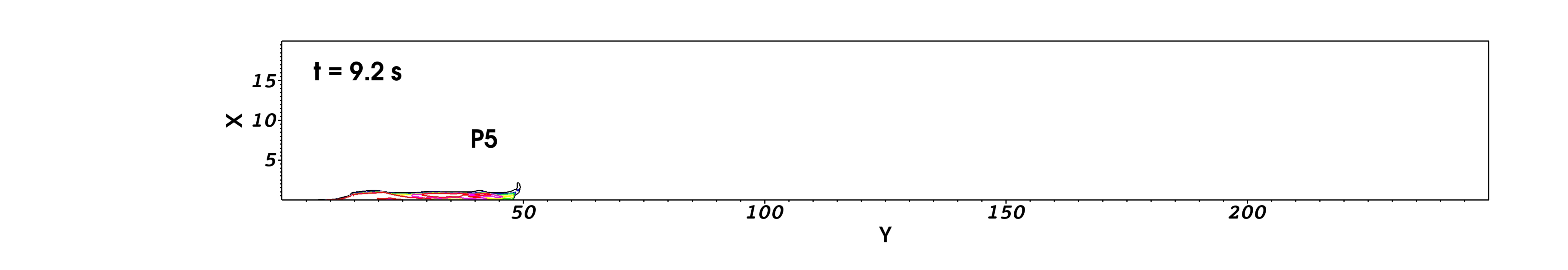}}  \\
 \subfloat{\includegraphics[width=0.9\linewidth]{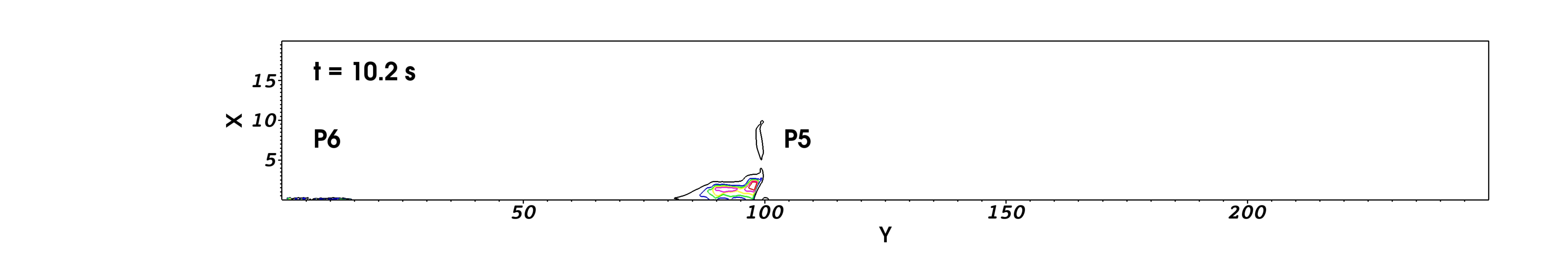}} \\
 \subfloat{\includegraphics[width=0.9\linewidth]{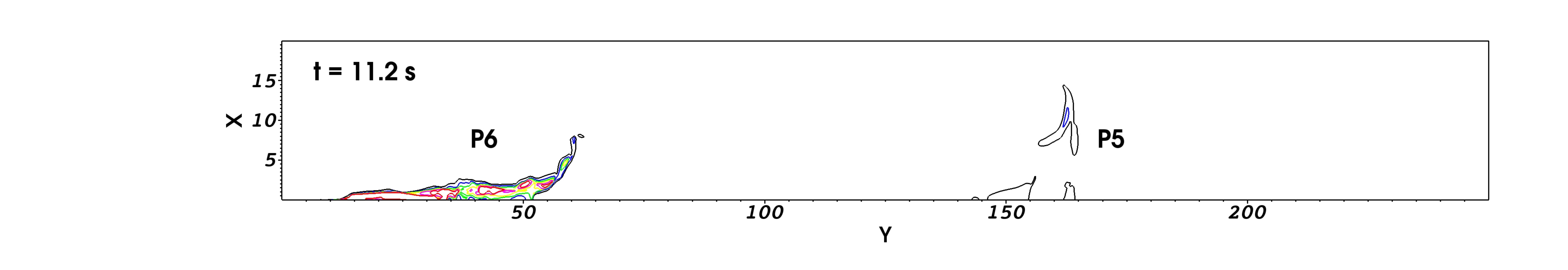}} \\
 \subfloat{\includegraphics[width=0.9\linewidth]{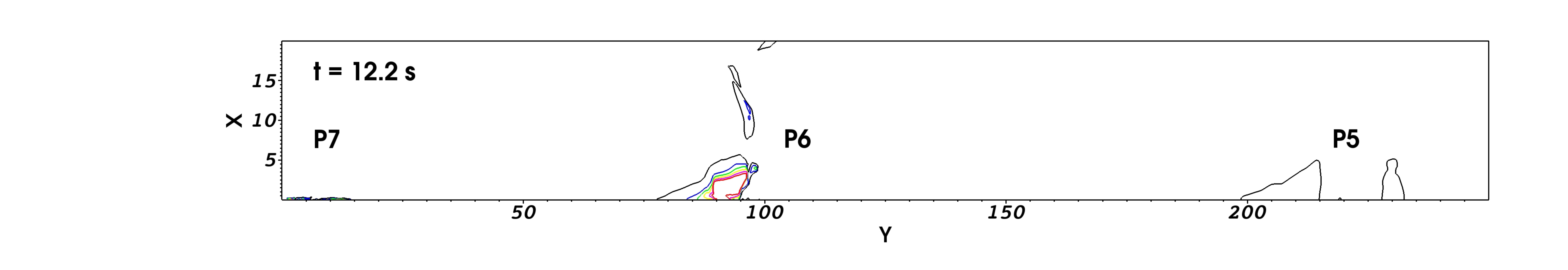}} \\
 \subfloat{\includegraphics[width=0.9\linewidth]{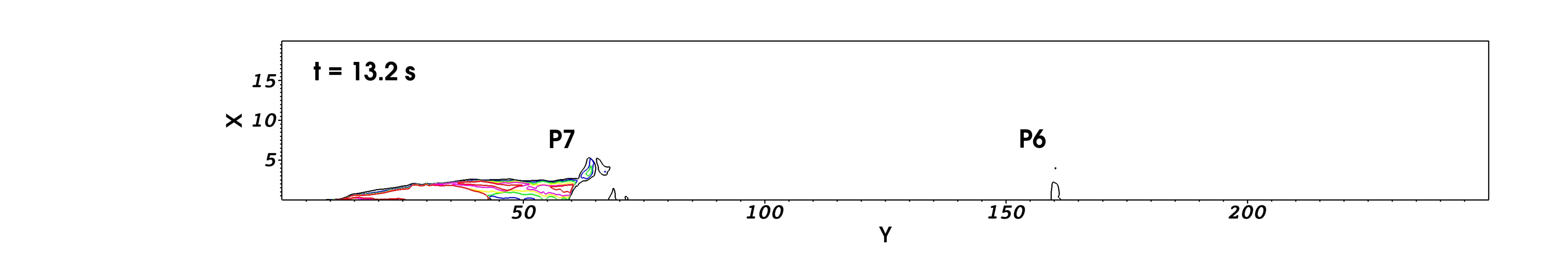}} \\
 \subfloat{\includegraphics[width=0.9\linewidth]{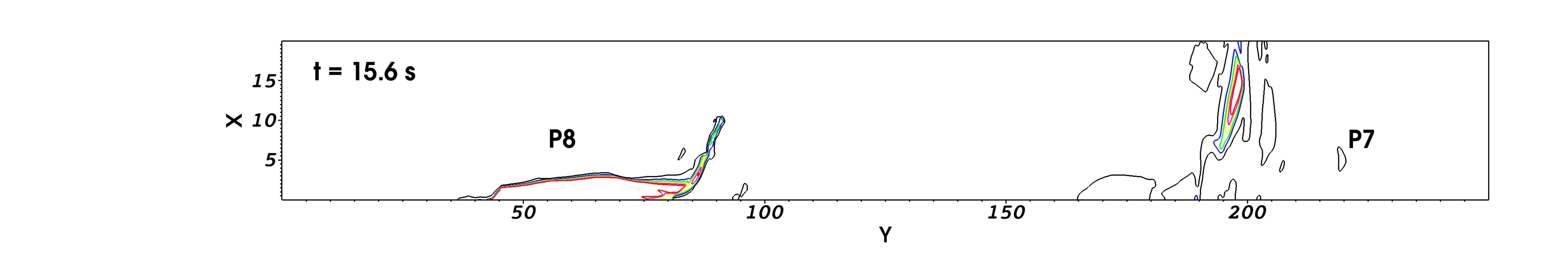}} 
\caption{Evolution of jet pulses ejected after $t_{\rm bo}$ in m10T1.0. The jet pulse
number \#N is denoted as ``PN''.
It can be seen that after the breakout time $t_{\rm bo}$ of the merged first three pulses
(P1-3), only the jet pulse ejected during 12-13 s
(P7) and the later ones can reach the outer boundary of the simulation box.
Other pulses (P4, P5 and P6) vanished on the way to the outer boundary.
The unit scale for X and Y axis is $4 \times 10^8$~cm.}
\label{Fig:plot9}
\end{figure}

\begin{figure}
   \begin{center}
   \includegraphics[scale=0.5]{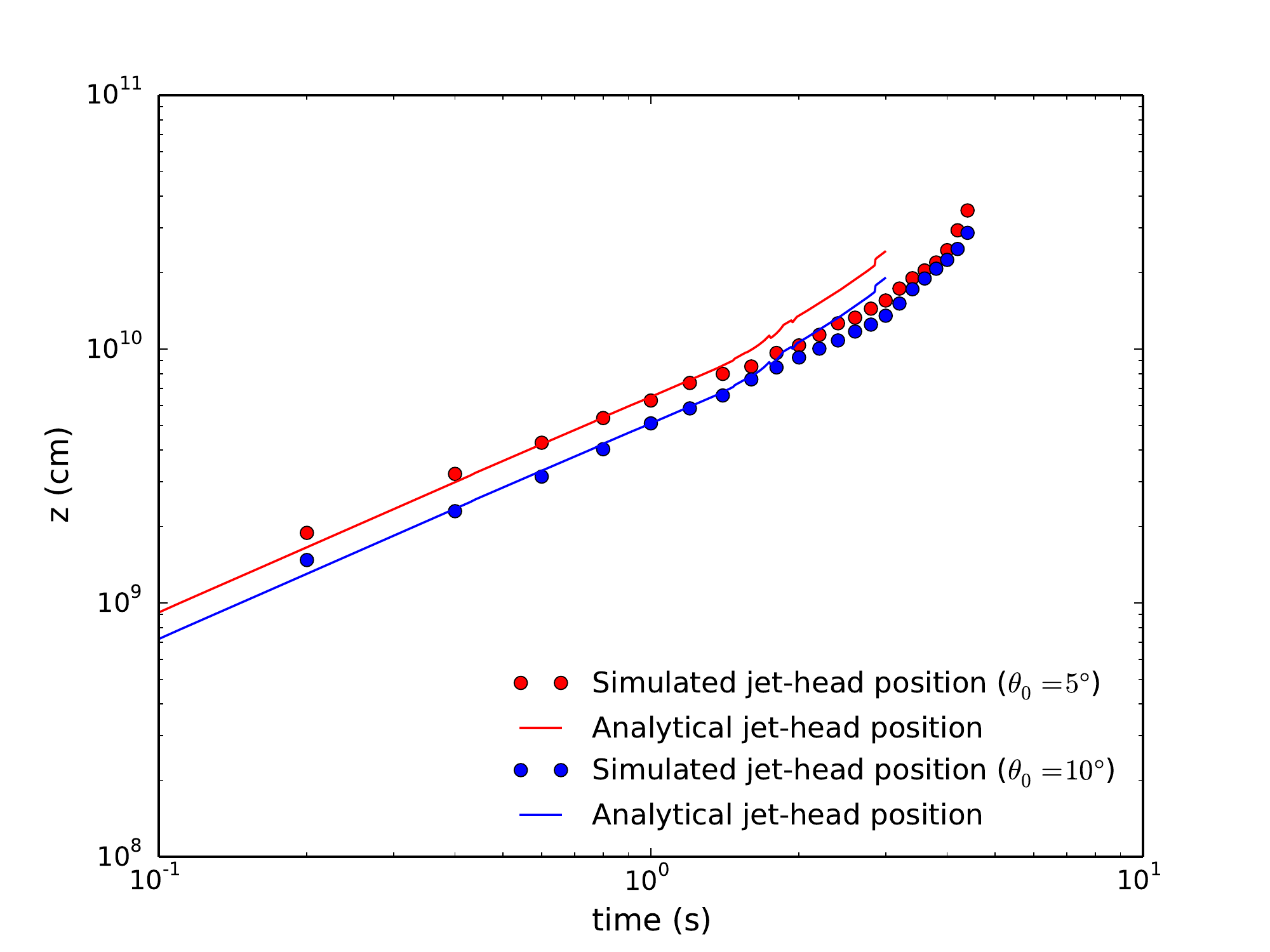}
   \caption{Comparison of the numerical results (dots) with the analytical results (solid lines) for the jet-head position.
   }
   \label{Fig:plot10}
   \end{center}
\end{figure}

\begin{figure}
   \begin{center}
   \includegraphics[scale=0.1]{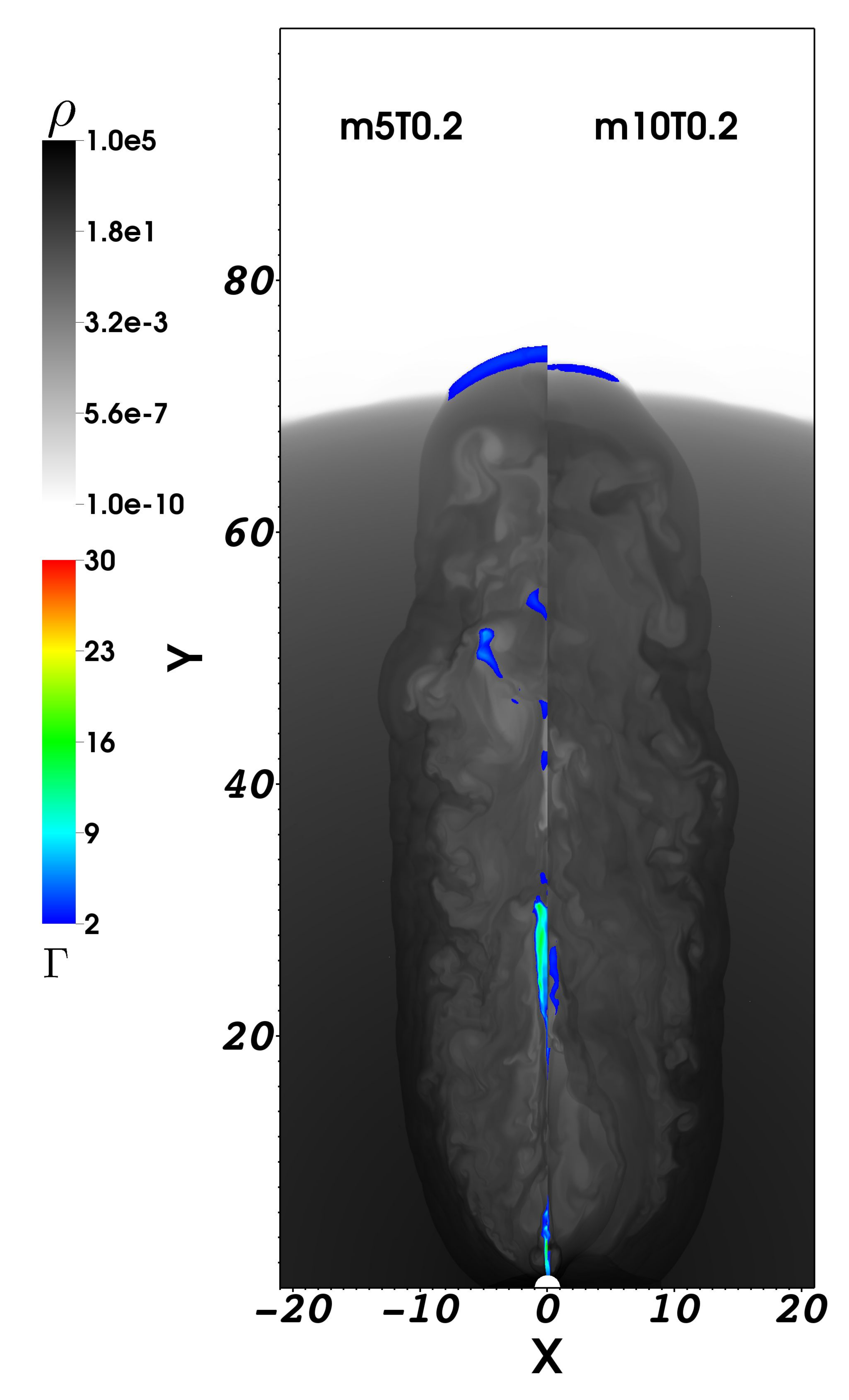}
   \caption{Comparison of Lorentz factor maps and density distributions between 
   m5T0.2 at $t = 4.9$ s (left) and m10T0.2 at $t = 6.5$ s (right) around the breakout time.
   The unit scale for X and Y axis is $4 \times 10^8$~cm.
   }
   \label{Fig:plot11}
   \end{center}
\end{figure}

\begin{figure}
   \begin{center}
   \includegraphics[scale=0.5]{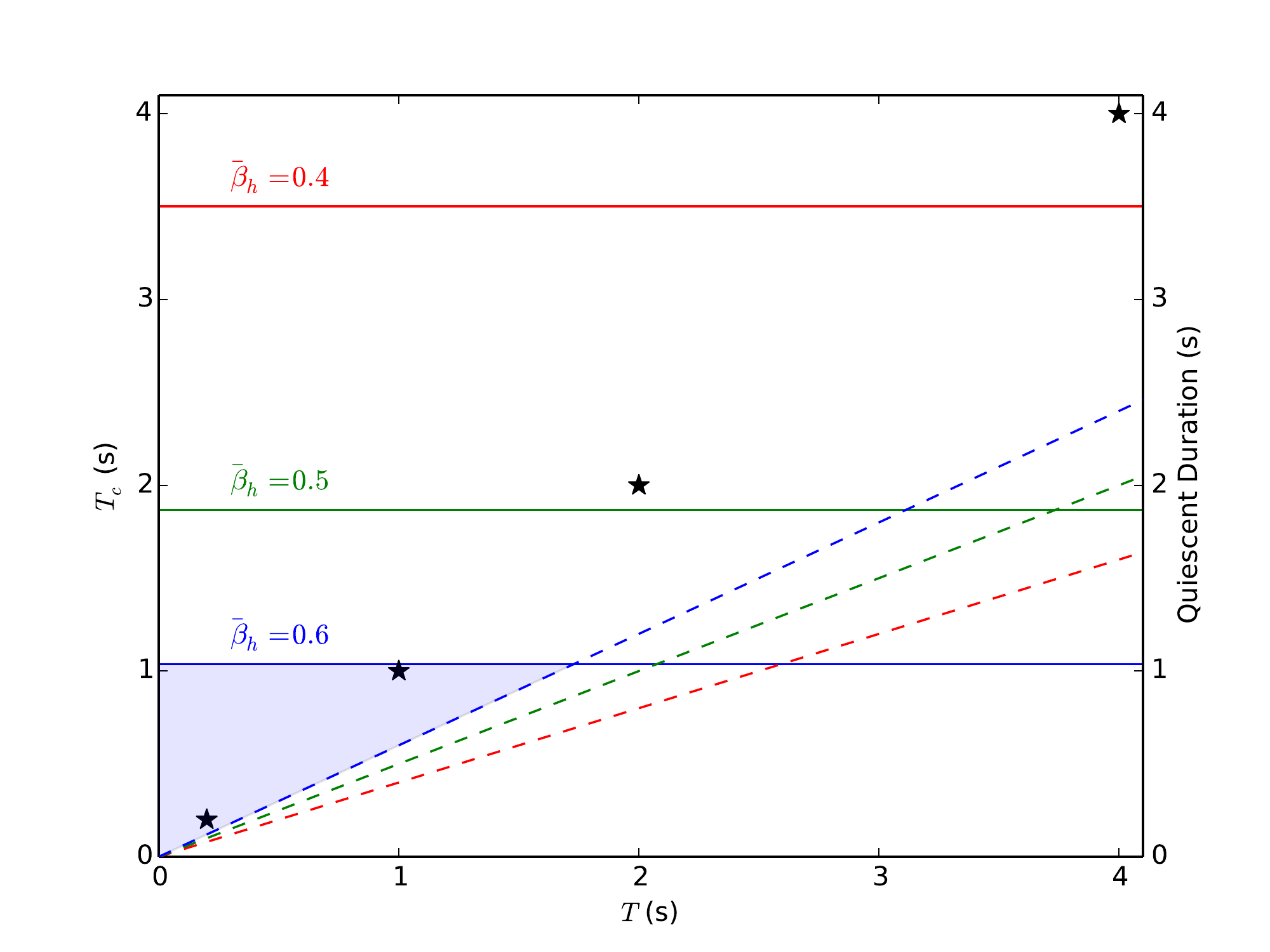}
   \caption{A schematic map of conditions for successful jet pulses.
   The solid lines show $T_c$ (Equation 13) by obtaining different $\bar{\beta}_h$.
   The dashed lines show the relationship between the quiescent time (above which the next jet 
   pulse would not catch up with the prior one) and $T$. 
   The colors indicate the different value of $\bar{\beta}_h$ used.
   Four simulation cases in this paper are marked by star symbols.
   It can be seen that the cases $T = 0.2$ s and $T = 1.0$ s are in the shaded region,
   within which the jet pulse vanishing would happen.
   }
   \label{Fig:plot12}
   \end{center}
\end{figure}

\begin{figure}
   \begin{center}
   \includegraphics[scale=0.1]{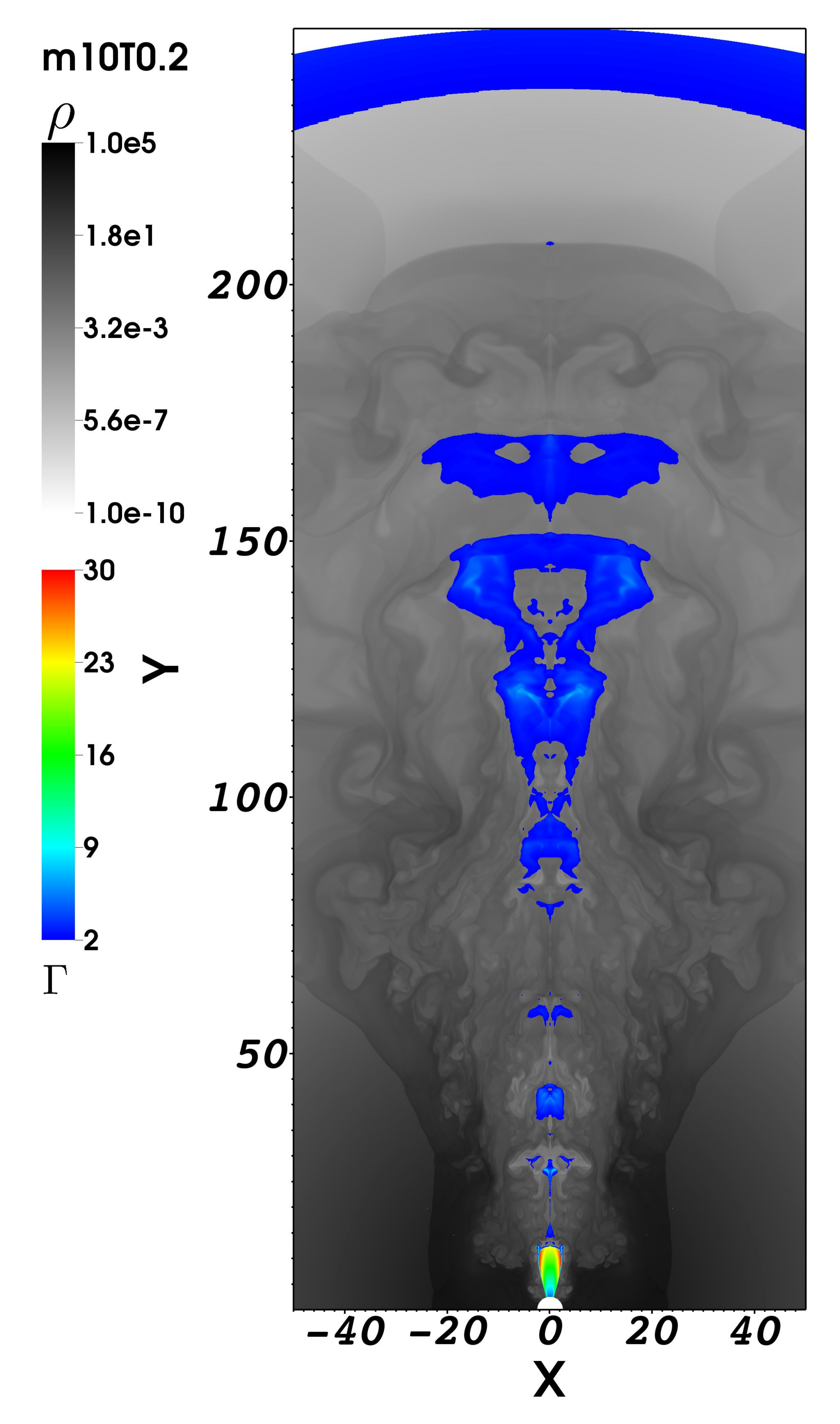}
   \caption{Lorentz factor map and density distribution (in units of g~cm$^{-3}$)
	of model m10T0.2 at $t = 9.8$ s, but the star envelope is not rotating or collapsing 
	(gravity is not considered).
	The unit scale for X and Y axis is $4 \times 10^8$~cm.
   }
   \label{Fig:plot13}
   \end{center}
\end{figure}

\begin{figure}
   \begin{center}
   \includegraphics[scale=0.1]{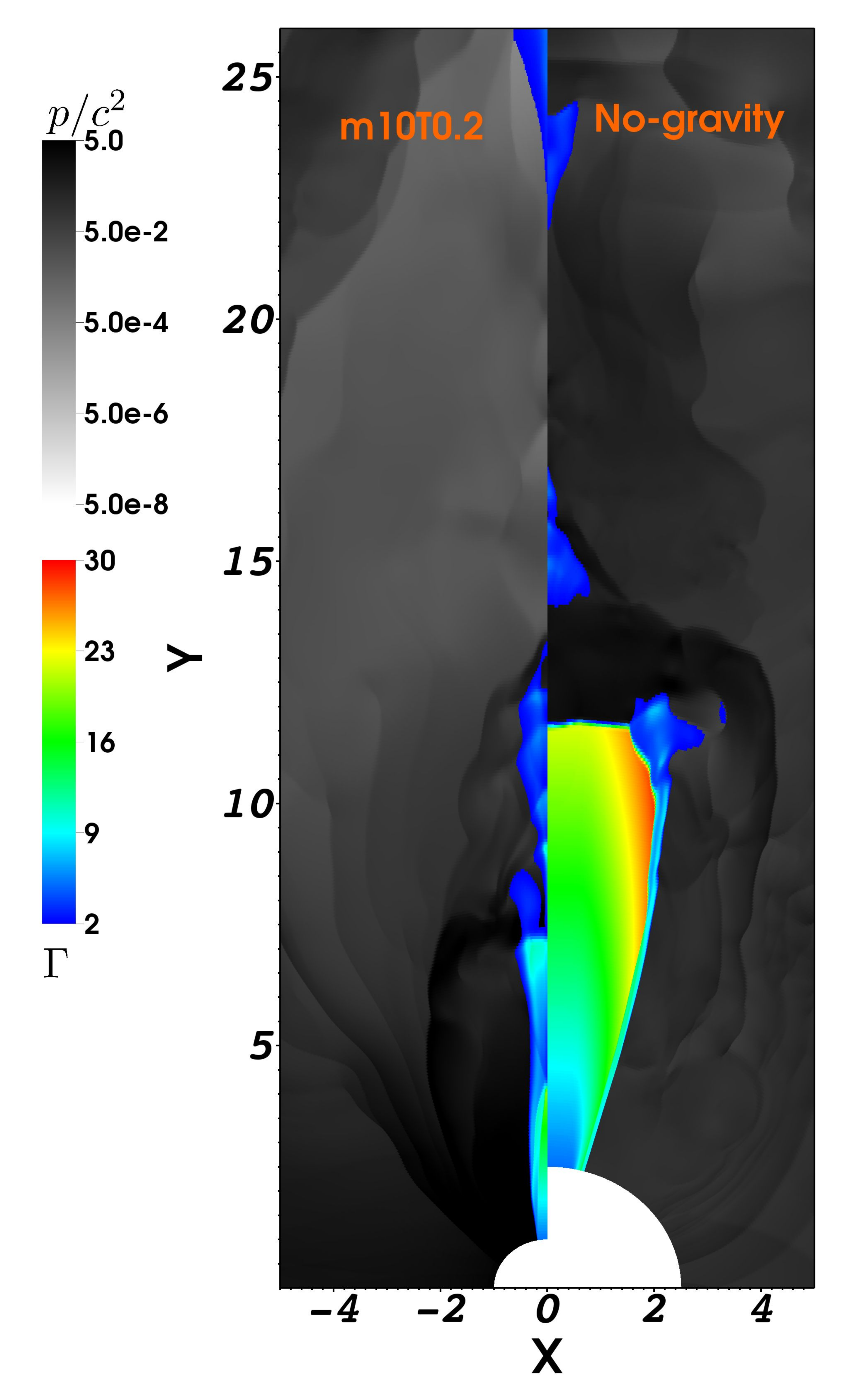}
   \caption{Comparison of Lorentz factor maps and pressure distributions between m10T0.2 
   with gravity (left) and m10T0.2 without gravity (right), both at $t = 9.0$ s. 
   The unit scale for X and Y axis is $4 \times 10^8$~cm.
   }
   \label{Fig:plot14}
   \end{center}
\end{figure}

\begin{figure}
   \begin{center}
   \includegraphics[scale=0.1]{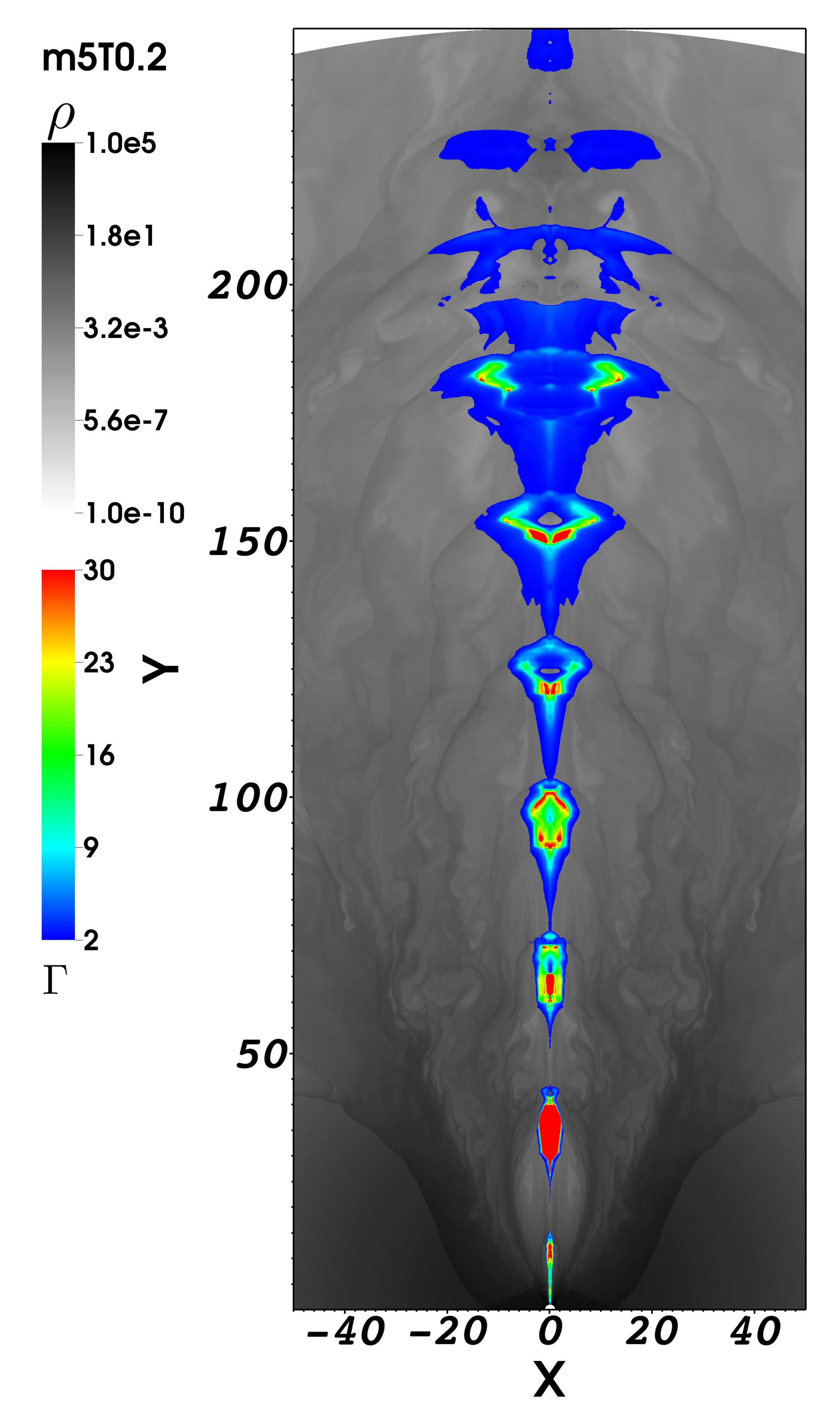}
   \caption{Lorentz factor map and density distribution (in units of g~cm$^{-3}$)
	of models m5T0.2 at $t = 11.4$ s. 
        The unit scale for X and Y axis is $4 \times 10^8$~cm.
   }
   \label{Fig:plot15}
   \end{center}
\end{figure}

\end{document}